\documentclass[floatfix,useAMS,usenatbib]{mn2e}
\usepackage[greek,british]{babel}
\usepackage{amsmath}
\usepackage{graphicx,epsfig}
\usepackage{multirow}
\usepackage{verbatim}
\usepackage[figuresright]{rotating}
\usepackage{txfonts}
\usepackage{times}
\usepackage[T1]{fontenc}
\usepackage{aecompl}

\def\msun{M$_{\odot}$}

\title[ Modeling the cometary structure of HFG1]
  {Modeling the cometary structure of the planetary nebula HFG1 based on the evolution of its binary  central star V664 Cas}
\author[Chiotellis et al.]
  {A. Chiotellis,$^1$\thanks{a.chiotellis@noa.gr}
  P. Boumis,$^1$ N. Nanouris,$^{1}$ J. Meaburn$^2$ 
  and G. Dimitriadis$^3$ \\
  $^1$Institute for Astronomy, Astrophysics, Space Applications
and Remote Sensing, National Observatory of Athens,
15236 Athens, Greece\\
 $^2$Jodrell Bank Centre for Astrophysics, University of Manchester, Manchester M13 9PL, UK\\
  $^3$School of Physics and Astronomy, University of Southampton, Southampton SO17 1BJ, UK}
\date{Released 2015 Xxxxx XX}

\begin{document}
\date{\ldots;\ldots}
\pagerange{\pageref{firstpage}--\pageref{lastpage}}\pubyear{2013}
\maketitle
\label{firstpage}

\begin{abstract}

HFG1 is the first well observed planetary nebula (PN) which reveals a cometary-like structure. Its main morphological features consist of a bow shaped shell, which surrounds the central star, accompanied by a long collimated tail. In this study we perform two-dimensional hydrodynamic simulations modeling the formation of HFG1 from the interaction of the local ambient medium with the mass outflows of its Asymptotic Giant Branch (AGB) progenitor star. We attribute the cometary appearance of HFG1 to the systemic motion of the PN with respect to the local ambient medium. Due to its vital importance, we re-estimate the distance of HFG1 by modeling the spectral energy distribution of its central star, V664 Cas, and we find a distance of  $ 490 \pm 50$~pc.  Our simulations show that none of our models with time invariant stellar wind and ambient medium properties are able to reproduce simultaneously the extended bow shock and the collimated tail observed in HFG1. Given this, we increase the complexity of our modeling considering that the stellar wind is time variable. The wind description is based on the predictions of the AGB and post-AGB evolution models. Testing a grid of models we find that the properties of HFG1 are best reproduced by the mass outflows of a 3 \msun~AGB star.  Such a scenario is   consistent with the current observed properties of  V664 Cas primary star, an O-type subdwarf,  and bridges  the evolutionary history of HFG1 central star with the observables of the PN.  We discuss the implications of our study in the understanding of the evolution of AGB/post-AGB stars towards the formation of O-type subdwarfs surrounded by PNe.

\end{abstract}


\section{Introduction}\label{sec:Intro}

Planetary nebulae (PNe) are large expanding shells formed by the mass outflows of  low/intermediate mass stars ($M \sim 1 - 8~\rm M_{\odot} $). These stars  during the asymptotic giant branch (AGB) phase  eject most of their envelope in the form of a strong stellar wind. Subsequently, the  fast wind that accompanies the contraction of the AGB core interacts with the previously ejected matter shaping the final morphology of the PN \citep{Paczynski71}. The resulting structure becomes visible as the gas is photoinonized by the hot central star - the remnant of the mass losing star - which is either a white dwarf (WD) or a subdwarf \citep{Kwok2000}.

\begin{figure}\label{fig:HFG1}
\includegraphics[trim=0 0 0 0,clip=true,width=\columnwidth,angle=0]{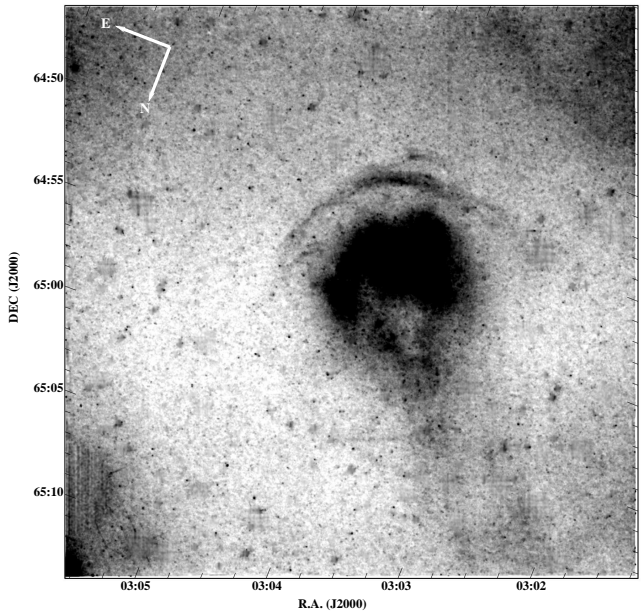} 
\caption {Combined $\rm H{\alpha}$ and [O {\sc iii}] data of the wide-field image of HFG1  [Credit: Ken Crawford - Rancho Del Sol Observatory].  The stellar contamination has been removed by implementing robust smoothing techniques based on the nearest pixel neighbors, while the distance to these neighbors (spread of each point source) has been adaptively adjusted to the local sky background  \citep[see also][for a more detailed description of the reduction pipeline]{Nanouris15}. The procedure has been applied both in the $\rm H{\alpha}$ and [O {\sc iii}] images and the combined image has emerged by co-adding the reduced images in 1:5 ratio.
  }
  \label{fig:nMdot}
\end{figure}

The PN HFG1,  { $ (\alpha,\delta)=~(03^{\rm h} 03^{\rm m} 47^{\rm s},~ 64^{\rm o} 54^{'}  35.7^{''}$)}, was first discovered  in  the deep  [O {\sc iii}] images  by \citet{Heckathorn82} using the emission line survey of \citet{Parker79}. Its morphology reveals a 9 arcmin diameter asymmetric nebula surrounded by a bow shaped outer ring of 15 arcmin diameter.  The brightness  of the outer ring  is also asymmetric, revealing its maximum value on the southeastern section of the nebula (Fig. \ref{fig:HFG1}).    The same authors performed spectroscopic studies of  the central region of HFG1 which show high excited plasma, a common feature of PNe. Finally, based on the $\rm H \beta$ surface brightness, they estimated  the distance of HFG1  to be 350 - 400 pc. Given this distance range, the radius of the nebula central region is 0.45 - 0.5 pc while this of the outer arc 0.75 - 0.9 pc. These properties classify HFG1 as a large, high excitation, double shell PN. 

\citet{Boumis09} obtained the first deep wide-field  $\rm H \alpha~+~$[N {\sc ii}]  ~6548~\&~ 6584 \AA~image of HFG1. In this image the nebula reveals a cometary structure where a tail of at least 20 arcmin long and 5 arcmin wide is lying in the opposite direction of the bow shock. The authors interpreted the cometary morphology of HFG1 to the interaction of the local interstellar medium (ISM) with the supersonically moving PN. They  reinforced their suggestion by measuring the proper motion (PM) of HFG1 and founding that the PN is moving with $\rm PM=13.0 \pm 1.5~ \rm mas~yr^{-1}$ with respect to the local ISM along a position angle of $\rm PA=133^o \pm 6^o$.

The central star (CS) of HFG1, named as V664 Cas, is  a binary system located in the core of the inner nebula  \citep{Heckathorn82}. Photometric variability of  V664 Cas, was first reported by \citet{Grauer87}, who  proposed that the  CS is a close detached pre-cataclysmic binary consisting of a hot primary which heats the  larger and cooler main-sequence companion.  Several multifilter photometric observations of V664 Cas \citep[{$V \sim 13.4$\,mag,}][]{Tylenda91} have been carried out since its discovery, suggesting a non-eclipsing binary with a light curve of a purely sinusoidal shape, {a wavelength-dependent amplitude ($\Delta m=$1.13, 1.14, and 1.12~mag in the V, R, I passbands, respectively)}, and a period equal to 0.58 days  \citep[e.g][]{Pigulski02}. The photometric period reflects the motion of the irradiated area (usually referred as a hot spot) of the secondary star - with respect to the line of sight - which is expected to be synchronized with the orbital motion, i.e. it is considered equal to the orbital period. The spectroscopic analyses of V664 Cas \citep{Shimanskii04,Exter05} classify the binary components as an O-type subdwarf (sdO, primary) and a F5-K0 main sequence star with roughly solar composition  (secondary).  Both \citet{Shimanskii04} and \citet{Exter05} modeled the V664 Cas spectra and light curves  providing a full set of fundamental parameters for the binary system.   Table \ref{tab:V664Cas_prop} shows the most reliable and well-constrained properties of V664~Cas as provided by  \citet{Shimanskii04}.   Finally, by estimating the color excess due to the interstellar absorption approximately equal to $E(B-V) = 0.5$ mag and assuming a spectral type range of F5-K0 V for the secondary, \citet{Exter05} have calculated a distance to HFG1 within a range of 310-950 pc.

\begin{table}
	\caption{ The fundamental parameters of the binary V664 Cas as estimated by \citet{Shimanskii04}. The primary and the secondary are labeled by the 1 and 2 subscripts, respectively. $M$ refers to  masses, $R$ to stellar radii, $g$ to  surface gravities and $T_{eff}$  to effective temperatures.}  
         	\begin{tabular}{|p{3.5cm}|p{3.5cm}| }
		\hline
	\hline
	Primary \\
	\hline
	$M_1~ \rm ( M_{\odot}) $ & $ 0.57 \pm 0.03 $ \\
	$R_1~  \rm ( R_{\odot}) $ & $0.19 \pm 0.02$ \\
	$\log g_1~ \rm (cm~s^{-2})$ &$ 5.65 \pm 0.05$ \\
	$T_{\rm eff, 1}~$(kK) & $83 \pm 6$\\
	\hline
	Secondary \\
	\hline
	$M_2~ \rm ( M_{\odot}) $ & $ 1.09 \pm 0.07 $ \\
	$R_2~  \rm ( R_{\odot}) $ & $1.30 \pm 0.08$ \\
	$\log g_2~ \rm (cm~s^{-2})$ &$ 4.23 \pm 0.06$ \\
	$T_{\rm eff, 2}~$(kK) & $5.4 \pm 0.5$\\
	\hline
	
		\end{tabular}
		\label{tab:V664Cas_prop}
\end{table}

All the aforementioned properties of HFG1 make this PN a very intriguing object as it shares two rare properties: a) only a few PNe with sdO central star   associations   have been reported in the literature  \citep[see table 5 of][and references therein] {Aller15}. The study of these PNe consists a vital tool to provide new insights into the unclear origin of sdO stars, and b) HFG1 is the only well observed PN that reveals a bow shock-tail cometary  structure. Detection of optical bow shocks is one of the cornerstones for determining, or at least constraining,  the properties of the stellar wind, the stellar proper motion, and the local ISM \citep{Gvaramadze12,Wareing12,Meyer14}. Therefore, HFG1 constitutes a unique celestial laboratory  which serves as a basis on a better understanding of the processes that accompany the end point  of low/intermediate mass stars evolution.  In particular, the bow shaped morphology of HFG1  provides  vital information on the mass loss history of its progenitor star as the mass loss rate and wind velocity can be contained by the geometrical properties of the PN.  This in turn provides new insights into the unknown evolutionary tracks toward the formation of sdO stars.

In this work, we demonstrate that the  characteristics of HFG1 can  be explained by the interaction of the local ISM with the stellar wind that was emanating during the AGB and post-AGB phase of its supersonically moving progenitor system. We perform 2D hydrodynamic simulations and we show that the cometary structure of the PN with a rather extended bow shock and a collimated tail  can best be reproduced by considering a time variable mass loss history as described by the theoretical predictions for the AGB wind of a 3~\msun~ star.   Such a result agrees with the observables of V664~Cas closing the loop from the HFG1 central star evolution to the morphological properties of the PN.

This paper is organized as follows: in Sect. \ref{sec:bow} we constrain the properties that determine the evolution of HFG1 based on the theory of bow-shaped  wind bubble formation.  In Sect. \ref{sec:distance} we re-estimate the distance of HFG1 eliminating its possible range. Based on that we further constrain the parameters which determine the model.  In Sect. \ref{sec:hydro} we model  HFG1 using hydrodynamic simulations. We describe the numerical code, the methodology  that we follow, and we present the results of our models. We evaluate our results and the implications of them in our understanding of PNe properties with sdO central star in Sect. \ref{sec:Disc}.  Finally, a summary of our conclusions is presented  in Sect.~\ref{sec:Sum}.

\section{ Formation of a bow shaped wind bubble}\label{sec:bow}

 Continuum mass outflows emanating from the stellar surface in the form of a stellar wind sweep-up the surrounding medium and form a circumstellar bubble. Initially, at the vicinity of the mass losing object, the  mass of the wind is much larger than this of the swept-up ambient medium (AM) and, consequently, the wind follows a free expansion phase. Within this region of the wind bubble the density drops off rapidly with increasing radius and it is given by: $\rho= \dot{M}/(4 \pi u_w r^2) $, where $ \dot{M}$, is the mass loss rate of the wind, $ u_w$ the wind terminal velocity  and $r$ the radial distance from the mass-losing object. In most cases the velocity of the stellar wind is supersonic with respect to the local AM. Therefore, a shock arises at the outermost part of the wind bubble  known as {\it forward shock}   which sweeps up the surrounding AM and compresses it in a dense, hot shell. The expansion of the wind bubble progresses, more and more swept-up mass is accumulated and the wind bubble starts to decelerate. The result of the deceleration is the formation of a second shock wave, the {\it termination shock}, which is propagating inwards, in the rest frame of the freely expanding region, and compresses the wind into a second shell. The two shells of shocked AM and wind material are separated by a contact discontinuity, of which the location is determined by the establishment of the pressure balance \citep[see][for a complete theoretical review]{Weaver77,Koo92}.

If the mass-losing object is moving supersonically, with respect to the local AM, the wind bubble is forming a bow shaped structure under the ram pressure of  the surrounding medium. If the AM is homogeneous, the stellar wind is spherically symmetric and time invariable  or at least its time variability is larger than the timescale of the wind outflow, 

\begin{equation}\label{eq:tflow}
t_{flow} = \left( \frac{r}{u_w} \right) \approx 10^5  \left(\frac {r}{pc} \right) \left( \frac{u_w}{10~km~s^{-1}}\right)^{-1}~ \rm yr ,
\end{equation}
with $r$ indicating the distance of the bow shock; the system relaxes in a {\it steady state}.  In this case, the structure of the bow shaped wind bubble  is determined by four variables: the wind's mass loss rate ($\dot{M}$) and terminal velocity ($u_w$), the spacial velocity of the mass losing object ($u_*$) and the AM density ($n_{AM}$).

The point of the bow-shaped termination shock closest to the mass losing star lies in the direction of the stellar motion and it is known as {\it stagnation point}. The radius of the stagnation point from the star ($r_0$) is estimated from the balance of the AM and stellar wind ram pressures ($P= \rho~u^2$) at the direction of the stellar motion.  This gives:
\begin{equation*}\label{eq:ramPress}
P_{wind}=P_{AM} \Rightarrow \frac{\dot{M}}{4 \pi u_w r_o^2}~u_w^2 = \rho_{AM}~u_*^2 \Rightarrow
\end{equation*}

\begin{equation*}\label{eq:ro}
r_0=   0.18  \left(\frac{\dot{M}}{10^{-5}~M_{\odot}~yr^{-1}} \right)^{1/2}   \left( \frac{u_w}{10~km~s^{-1}} \right)^{1/2}
\end{equation*}
\begin{equation}\label{eq:stag}
~~~~~~~~~~~~~~~~~~~~~~\times \left( \frac{n_{AM}}{cm^{-3}} \right)^{-1/2} \left( \frac{u_*}{100~km~s^{-1}} \right)^{-1}  ~\rm pc.
\end{equation}

The approximate shape of the bow shaped termination shock with respect to the stagnation point is given by: 

\begin{equation}\label{eq:r/ro}
\frac{r}{r_0}=  \frac{\theta}{\sin \theta},
\end{equation}
where $r$ is the distance between the mass losing object and the termination shock, while $\theta$ is the angle between the radius vector $r$ and the vector of the stellar velocity  $u_*$  \citep{Houpis80,Borkowski92}. In the opposite direction of the stellar motion, the shell of the shocked wind is compressed by the AM ram pressure and for the cases where the later is much higher than that of the wind pressure, the socked wind shell is getting accumulated in a narrow region behind the mass losing object, which observationally resembles a `tail'.

As  described in Sect. \ref{sec:Intro},  the PN HFG1 reveals a bow shock-tail morphology  which  most likely is formed by the interaction of the local AM with the wind of its supersonically moving progenitor. \citet{Boumis09} estimated the spacial velocity of the PN, with respect to the local AM, to be in the range from $u_*= 29 \pm 4~ km~s^{-1}$ to $ 59 \pm 9~ km~s^{-1}$ for the given range of possible distances of $D= 310 -950$ pc \citep{Exter05}, respectively.  The angular radius of HFG1 stagnation point (radius of the outer arc at the direction of motion) is $\sim 6.5$~arcmin \citep[see fig.2 of][]{Heckathorn82}, which corresponds to a stagnation point radius of $r_0= 0.59 - 1.80$~pc for the aforementioned distance range.

To reproduce the radius of the stagnation point, except from the known - distance dependent - spacial velocity of HFG1, we also need  the wind properties ($\dot{M}, u_w$) of its progenitor as well as the density of the local AM. As  described in the Introduction, PNe are formed by the mass outflows at the ending phase of AGB stars. These stars are characterized by strong, slow stellar winds with mass loss rates of $\dot{M} \approx 10^{-7} - 10^{-5}~ \rm M_{\odot}~yr^{-1}$ and wind terminal velocities of $u_w \approx 5 - 15~ \rm km~s^{-1}$, depending on their mass, metallicity and AGB evolutionary stage \citep{Vassiliadis93}. As far as the local AM density is concerned, we do not have any a priori estimates. However, its range can be estimated by the constraint of the observed value of the stagnation point radius given the aforementioned range of PN spatial velocity, the wind mass loss rate and the wind velocity (see Eq. \ref{eq:ro}). This implies that the number density should be in the range of $n_{AM}= 5\times 10^{-3} - 1.6~ \rm cm^{-3}$ and $n_{AM}= 5\times 10^{-4} - 4 \times 10^{-2} \rm~ cm^{-3}$ for the two limits of the possible distance range, $D= 310$~pc and $D= 950$~pc, respectively (Fig. \ref{fig:nMdot}). Such AM density values are reasonable as they are characteristic for the neutral/ionized warm or hot component of the ISM (see e.g. McKee \& Ostriker 1977). Finally, for the given radius of the stagnation point and assuming the typical values of AGB winds, a steady state situation is reached after $t_{flow} \approx 0.05 - 0.1$~Myr for  $r_0= 0.59 $~pc and  $t_{flow} \approx 0.2 - 0.3$~Myr for $r_0= 1.80 $~pc. These timescales are well within the limits of the lifetime of AGB stars ($t_{AGB}\sim 1$~Myr).

\begin{figure}
\includegraphics[trim=0 0 0 0,clip=true,width=\columnwidth,angle=0]{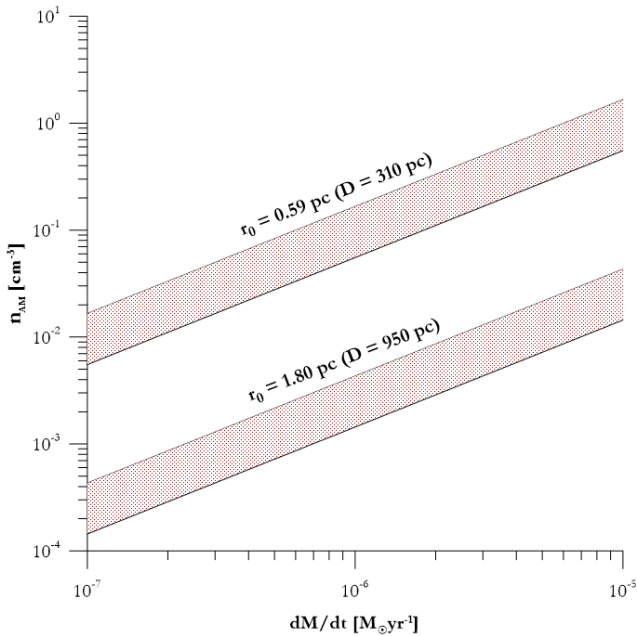} 
\caption { The AM density versus the characteristic mass loss rates of AGB stars for values satisfying the observed stagnation point radius of HFG1 (see Eq. \ref{eq:ro}). The two shadowed strips correspond to the extreme limits of possible HFG1 distances. The width of each strip reflects the AGB wind velocity range: $u_w \approx 5 - 15~ \rm km~s^{-1}$. }
  \label{fig:nMdot}
\end{figure}

\section{ Eliminating the distance range of HFG1}\label{sec:distance}

The wide range of HFG1 possible distances estimated by \citet{Exter05} introduces large uncertainties to the model as both the PN systemic velocity and the size of its structure are distance dependent (see previous section).

In this work, attempting to  eliminate the distance range, we proceed to an updated distance estimation by implementing a modified approach of \citet{Bonanos06, Bonanos11}. The procedure aims to model the observed spectral energy distribution (SED) $f_{\lambda}$ of the PN central binary system based on the optimal set of physical parameters that \citet{Shimanskii04} provide for its stellar components.

      To realize this task, phased apparent magnitudes of V664~Cas in a wide range of passbands are required; however, no such information is available in the literature due to the lack of published minima timings and light measurements converted to the standard photometric systems. To deal with this incompleteness, we first phase the available magnitudes at the near infrared wavelengths, as provided from the 2MASS photometric survey (Cutri et al. 2003), based on the seasonal ephemeris of \citet{Shimanskii04}; since the latter was composed a few only months later than the epoch the 2MASS measurements were conducted, it is accurate enough for this purpose. We find that the values of $J = 12.93 \pm 0.02$ mag, $H = 12.68 \pm 0.03$ mag, and $K_s = 12.58 \pm 0.03$ mag actually correspond to the zero phase, i.e., to the minimum of the light curve where the non-irradiated side of the secondary component is observed. Intending to complement with the visual counterparts, an intensive program of observations was implemented in the visual passbands at two observatories, as thoroughly described in the following section.

\subsection{Observations}\label{subsec:obs}

 \begin{figure*}
\begin{center}$
\begin{array}{cc}

 \includegraphics[trim=0 0 0 0,clip=true,width=85.0mm,angle=0]{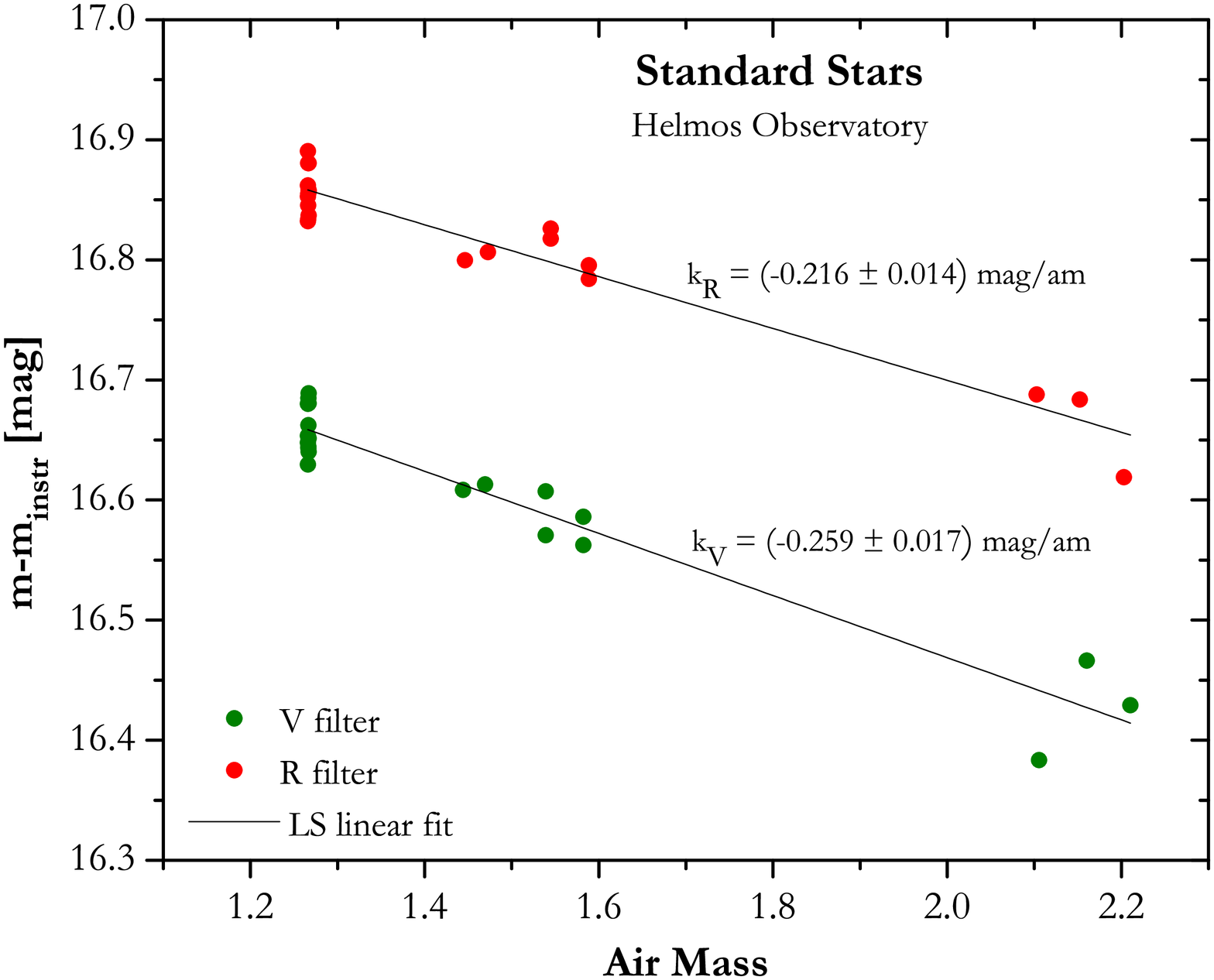} &
\includegraphics[trim=0 0 0 0,clip=true,width=87.0mm,angle=0]{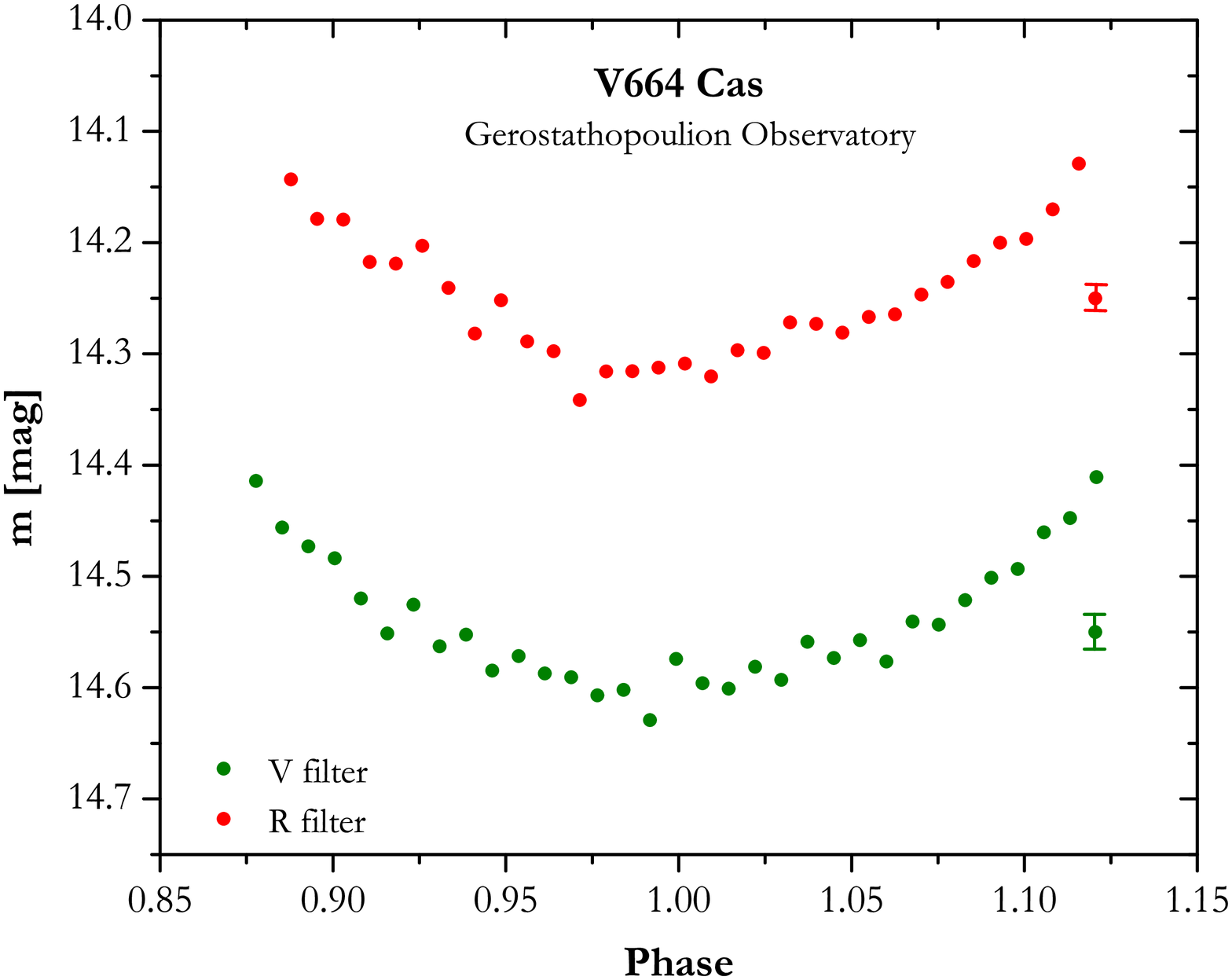} \\
\end{array}$
\end{center}
\caption{ Left: The linear atmospheric extinction curve of the monitored standard stars as emerged from photometric measurements at Helmos site on August 27, 2015. The first order extinction coefficients with their standard errors are also given for the V (green dots) and R (red dots) passbands. Right: Photometric observations of V664 Cas at the V (green dots) and R (red dots) passbands with a mean standard error of 0.015 mag. The measurements have been converted to the standard Johnson-Cousins photometric system after the appropriate atmospheric extinction corrections.}
  \label{fig:standarstars}
\end{figure*}

      Photometric observations of standard stars took place with the 2.3~m , Aristarchos telescope (f/8) at  Helmos observatory on August 27, 2015 with very good atmospheric conditions (seeing $\sim 1.4$ arcsec).  It is equipped with a liquid nitrogen CCD camera of a $1024 \times 1024$ pixel$^2$ array and a $5 \times 5~\rm arcmin^2$ field of view, resulting in a $0.29~\rm arcsec^2$ pixel resolution. Eight standard stars were properly selected from the \citet{Landolt92} catalogue to be monitored in a large range of air masses (94401, 110266, 110360, 110361, 110364, 113163, 113167, and 114531). The photometric measurements were acquired at the V, R broad Johnson-Cousins passbands with exposures varying from 5 to 60 seconds, depending on the filter and the program star, achieving an error close to or less than 5 mmag. The extinction coefficient and the instrumental offset were then estimated through a standard least-squares linear fit (Fig. \ref{fig:standarstars}, left). No attempt was made to correct for the color effects; that is, the second order extinction coefficient was omitted from the overall analysis. In addition, differential photometry of V664 Cas was carried out on August 28, 2015 covering a small part of the light curve, however other than the minimum profile. GSC 04056:01235 ($ \alpha= 03^h 03^m 54^s, \delta = + 64^o 55' 58''$) was used as a comparison star, while N313330017403 ($\alpha = 03^h 03^m 41^s, \delta = + 64^o 56' 07''$) confirmed the photometric constancy of the latter.

Additional observations were performed at the 0.4 m telescope (f/8) of the Gerostathopoulion observatory at the University of Athens on September 16, 17, 18 and 19, 2015. The observatory is equipped with a thermoelectrically Peltier - cooled CCD camera of a $2184 \times 1472$ pixel$^2$ array and a $16 \times11~\rm arcmin^2$ field of view, resulting in a $0.45\rm arsec^2$ pixel resolution. GSC 04056:01235 was still preferred as comparison to facilitate the precise determination of the desired V, R magnitudes for V664~Cas. A mean error close to 0.015 mag was reached with exposures of 120 sec. GSC~04056:00045 ($\alpha = 03^h 04^m 31^s, \delta= + 64^o 55' 23''$) was selected as the check star; no variations were detected in their residuals during the four-nights monitoring. By calculating two minima timings, a new ephemeris was constructed to phase all available observations.    

 All data accounting for the differential photometry were reduced using the software MuniWin v.1.1.28 \citep{Hroch98}, while the absolute photometry was performed through the appropriate IRAF routines \citep{Tody93} in the IDL programming environment. The magnitude values were obtained by the standard aperture photometric procedure, while the observing times were converted to the heliocentric Julian Date (HJD).

 In the absence of an accurate ephemeris during the observing run at  Helmos observatory, the observed fluxes of the standard stars were employed to calibrate the comparison star GSC 04056:01235, yielding the values $V_{comp} = 13.23 \pm 0.03$ mag and $R_{comp} = 12.69 \pm 0.03$ mag. The more extensive differential photometry at the Gerostathopoulion observatory then allowed the precise estimation of the V, R magnitudes for V664~Cas at the minimum of the light curve (Fig. \ref{fig:standarstars}, right). More particularly, the values $V = 14.60 \pm 0.04$ mag and $R = 14.31 \pm 0.03$ mag were finally inferred from the analysis, taking also into account a small offset between the two observatories (by comparing the differential photometry at the same phases).

\subsection{ Mathematical procedure}\label{subject:mathproc}

 To convert our visual and 2MASS near infrared magnitudes to fluxes, we use zero points from \citet{Bessell98} and \citet{Cohen03} as explicitly described in \citet{Bonanos06, Bonanos11}. Because of the multi-parameter nature and the mathematical complexity of the physical problem, we assume that the surface fluxes of both components are well reproduced by a black-body Planck distribution; since the observed SED corresponds to the minimum of the light curve, this representation holds for the non-irradiated only side of the cool component which is the less contaminated from the reflection effect. To further simplify the mathematical approach and to guarantee convergence, we then remove the impact of the primary component from the system, taking into consideration the range in which the hot subdwarf absolute parameters vary ($T_{eff,1}=~83000 \pm 6000$ K, $R_{1} = 0.19 \pm 0.02 \rm ~R_{\odot}$, see Table \ref{tab:V664Cas_prop}).

\begin{figure}
\includegraphics[trim=0 0 0 0,clip=true,width=\columnwidth,angle=0]{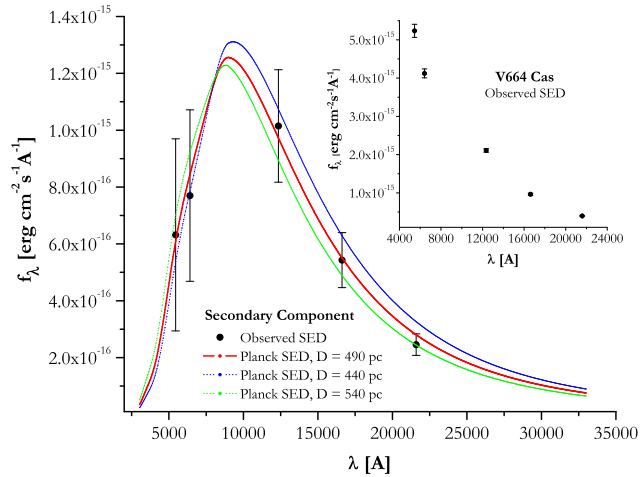} 
\caption {  The observed spectral energy distribution (SED) over the available VRJHKs photometric bands for the binary system V664 Cas (see enclosed plot) and its cool only component. The regression curves correspond to a Planck distribution of 5400 K, as emerged from the non least-squares fitting procedure. The 490 pc optimal model is depicted with the red solid line, while the $1\sigma$ lower and upper distance of 440 and 540 pc are displayed with the blue and green dashed lines, respectively. }
  \label{fig:distance}
\end{figure}

 In the remaining SED of the secondary component, the observed fluxes $f_{\lambda}$ are accompanied by their formal errors reflecting both the observational errors and all the uncertainties caused from the primary SED subtraction over all available $\rm VRJHK_s$ photometric bands (Fig. \ref{fig:distance}). Given the Planck-reproduced surface fluxes $F_{\lambda}$ and the radius $R$ of the main sequence component, the distance $D$ can be then made available through the following relation:

\begin{equation}\label{eq:Fl}
f_{\lambda}= \frac{1}{D^2}~R^2~F_{\lambda}~10^{-0.4 A_{\lambda}},
\end{equation}
with the extinction curve $A_{\lambda}$  being calculated as a function of the color excess $E(B-V)$ via the reddening parameterization of \citet{Cardelli89}, adapting both the optical and near infrared wavelengths  and setting $R_v= A_v/E(B-V) =3.1$ as reference value for the visual. The distance is then determined by means of a non-linear least-squares minimization Gauss-Newton procedure \citep[e.g.][]{Press92}. To involve the uncertainties of the observed SED, the fluxes are weighted proportionally to the inverse value of their standard errors.

In its simplest version, the aforementioned regression scheme is first implemented by fixing the effective temperature, the stellar radius, and the color excess at 5400 K, $\rm 1.3 R_{\odot}$ \citep{Shimanskii04}, and 0.5 mag \citep{Exter05}, respectively, { leading to a distance $640 \pm 50$ pc.} In a more advanced approach, the color excess is set as an unknown parameter, keeping the same temperature and radius values fixed as before. { A shorter distance of $497 \pm 15$~pc and a stronger color excess of $0.83 \pm 0.04$ mag} are then derived as the optimal pair after the final convergence.

However, the inferred errors seem to be unreliably small since they do not reflect the uncertainties of all the incorporated physical quantities.  To raise the validity of the approach, we take into consideration both the range in which the stellar absolute parameters vary \citep[$T_{eff,2} = 5400 \pm 500$~K, $R_2 = 1.30 \pm 0.08~\rm R_{\odot}$,][] {Shimanskii04}, and the residuals scatter (as emerged from the best-fit model) by performing Monte Carlo (MC) simulations through suitable R routines \citep{Rcore}. In particular, we generate (a) 1000 temperature values from a Gaussian distribution with mean 5400 K and standard error 500 K, (b) 1000 radius values from a Gaussian distribution with mean 1.3 $\rm R_{\odot}$ and standard error 0.08 $\rm R_{\odot}$, and (c) 1000 synthetic residual samples following a zero-centered Student distribution with 3 degrees of freedom and spreading by $6.9 \times 10^{-17} \rm erg~cm^{-2}~s^{-1}~A^{-1}$, which corresponds to the residual mean error of the 5400 K optimal model. Note that in the presence of only a few available data, a t-distribution is a more appropriate choice instead of a Gaussian to model the residuals. The simulated values yield a distance $D = 490 \pm 50$ pc and a color excess $E(B-V) = 0.82 \pm 0.14$ mag interval, respectively, with their upper and lower limits to represent any uncertainties may be propagated from the physical problem parameter space at a 68\% significance level. The MC error was estimated as low as 5 pc and 0.02 mag (after having repeated the same procedure by 10 times), confirming that the sampling was sufficient enough for inference.

Although all aforementioned approaches give distance values  which are in good agreement with those found in  literature, here we have managed to eliminate the range in which the distance lies. The   larger color excess values with respect to those inferred from spectrophotometric analyses \citep{Heckathorn82, Exter05} imply that the interstellar extinction might have been underestimated so far, a finding that is consistent with the value of $0.92 \pm 0.03$ mag, as arisen from the Galactic reddening mapping of Schlafly \& Finkbeiner (2011) in the nearby region. Similarly to \citet{Exter05}, we have isolated the SED of the secondary component; however, instead of the red only passband, we have taken advantage of five available photometric bands. Even further, taking the variability of V664 Cas into consideration, the uncertainties of both the full physical parameter set and the regression approach, we suggest a more well-constrained distance of $490 \pm 50$ pc.

Adopting a distance of $D= 490$~pc for HFG1  the observed angular distance of the outer shell stagnation point corresponds to a radius of $r_0= 0.92$~pc while its spacial velocity with respect to the local AM is $u_*= 35 \pm 5~ km~s^{-1}$.  Following the process of Sect.~\ref{sec:bow} we find that, for the AGB wind properties, an AM density of  $n_{AM}= 1.5\times 10^{-3} - 0.5~ \rm cm^{-3}$ is needed to reproduce the stagnation point, while a steady state situation is reached for $t_{flow} \approx 0.1$~Myr.

\section{Hydrodynamic modeling}\label{sec:hydro}

In Sect. \ref{sec:bow}  we show that the scenario which suggests that the cometary structure of HFG1 is formed by the mass outflow of its supersonically moving AGB progenitor seems able to reproduce the observed position of the stagnation point adopting characteristic values of AGB winds  and AM densities which correspond to the  neutral/ionized warm or hot component of the ISM. Here, we further investigate this scenario by performing two dimensional (2D) hydrodynamic simulations,  targeted to model the overall morphology of HFG1.

\subsection{Method}\label{sec:Method}

We employ the hydrodynamic code of the {\sc amrvac} framework \citep{Keppens03} to simulate  the formation and evolution of the PN HFG1 that surrounds the pre-cataclysmic binary V664 Cas. We perform the computations on a 2D grid in spherical coordinates  assuming symmetry in the third dimension. The Euler equations are solved conservatively with a TVDLF  scheme, using the adaptive mesh strategy to refine the grid where needed as a result of large gradients in density and/or energy. Our radial span is $35\times10^{18}$~cm and the range of the polar angle is from  $0^\circ$ to $180^\circ$. On the base level, we use $R \times \theta$ = $96 \times 60$ cells  and  allow for three refinement levels, at each of which the resolution is doubled. The maximum effective resolution, thus, becomes  $9.1 \times 10^{16}$~cm by $0.75^\circ$.  Radiative cooling is prescribed using the cooling curve of \citet{Schure09}.

 We model the system in the rest frame of the PN progenitor and we represent the ISM interaction as an inflow. The ISM of density $\rho_{AM}$ enters the grid antiparallel to the y-axis and with a momentum $m_r=\rho_{AM} u_{*} \cos \theta$. Thus, the symmetry axis is aligned with the systemic direction of motion.
In the inner radial boundary, we impose a continuous inflow in the form of a stellar
wind with a density profile of $\rho = \dot{M}_w/(4\pi r^2 u_w )$ and momentum components $m_r = \rho u_w$ and $m_{\theta} =0$.

\begin{figure}
\includegraphics[trim=0 0 0 0,clip=true,width=\columnwidth,angle=0]{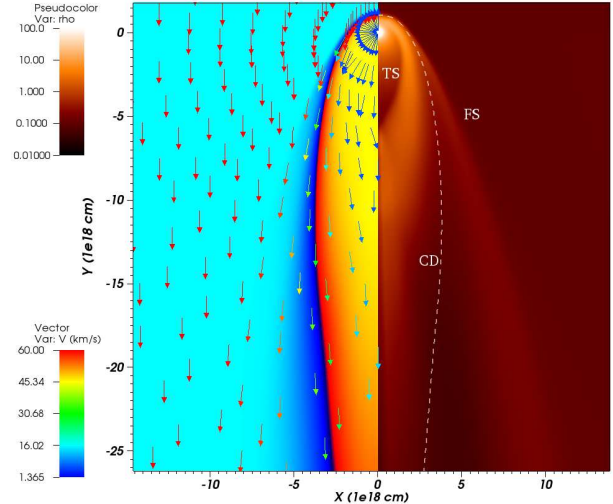} 
\caption { The 2D profile of a bow shaped wind bubble. The  stellar wind emanates from the axis origin where we impose a radial spherically symmetric flow to enter the grid.  The motion of the AM in the star's rest frame is represented by a parallel, homogeneous flow that simultaneously enters the grid in the direction of the y-axis. The right plot shows the 2D density profile while the left displays the region of the wind material (yellow) and this of the AM (cyan). The red and dark blue colors represent the regions where these two flows are mixed. The arrows correspond to the velocity vectors of each flow.  Finally, the symbols ÔFSÕ, ÔCDÕ, ÔTSÕ, mark the positions of the forward shock, contact discontinuity and termination shock, respectively.     }
  \label{fig:hydrosample}
\end{figure}

Figure \ref{fig:hydrosample} illustrates a typical structure of a wind bubble, formed by a supersonically moving mass losing star. From inside out are clearly depicted the four regions of the freely expanding wind where $\rho \propto r^{-2}$, the shock wind shell, the shell of shock AM, and the region of unperturbed AM (see right plot of the figure). The inner density jump corresponds to the position of the termination shock,  while the outer one to this of the forward shock. The dashed line marks the position of the contact discontinuity which separates the wind material from the AM.

\subsection{Models with constant ambient medium and wind properties}

In Sect. \ref{sec:bow}, we defined the possible range of the AGB wind properties  as well as the constraints of the AM density,  placed by the  observed - distance dependent - radius of the stagnation point, and the systemic velocity of the central star. These ranges/constraints combined with the observed systemic velocity of HFG1 determine  the input parameter space of our models. Taking this into account  we perform a grid of hydrodynamic models attempting to find the optimal parameters that are able to reproduce the main characteristics of HFG1.

  \begin{figure*}
\begin{center}$
\begin{array}{cccc}
&\rm \bf Low~AM/wind~ram~pressure~ratio & \rm \bf High~AM/wind~ram~pressure~ratio &
\\
\bf D = 310~pc: 
\\
\bf D= 950~pc: & \includegraphics[trim=15 50 85 20,clip=true,width=70.0mm,angle=0]{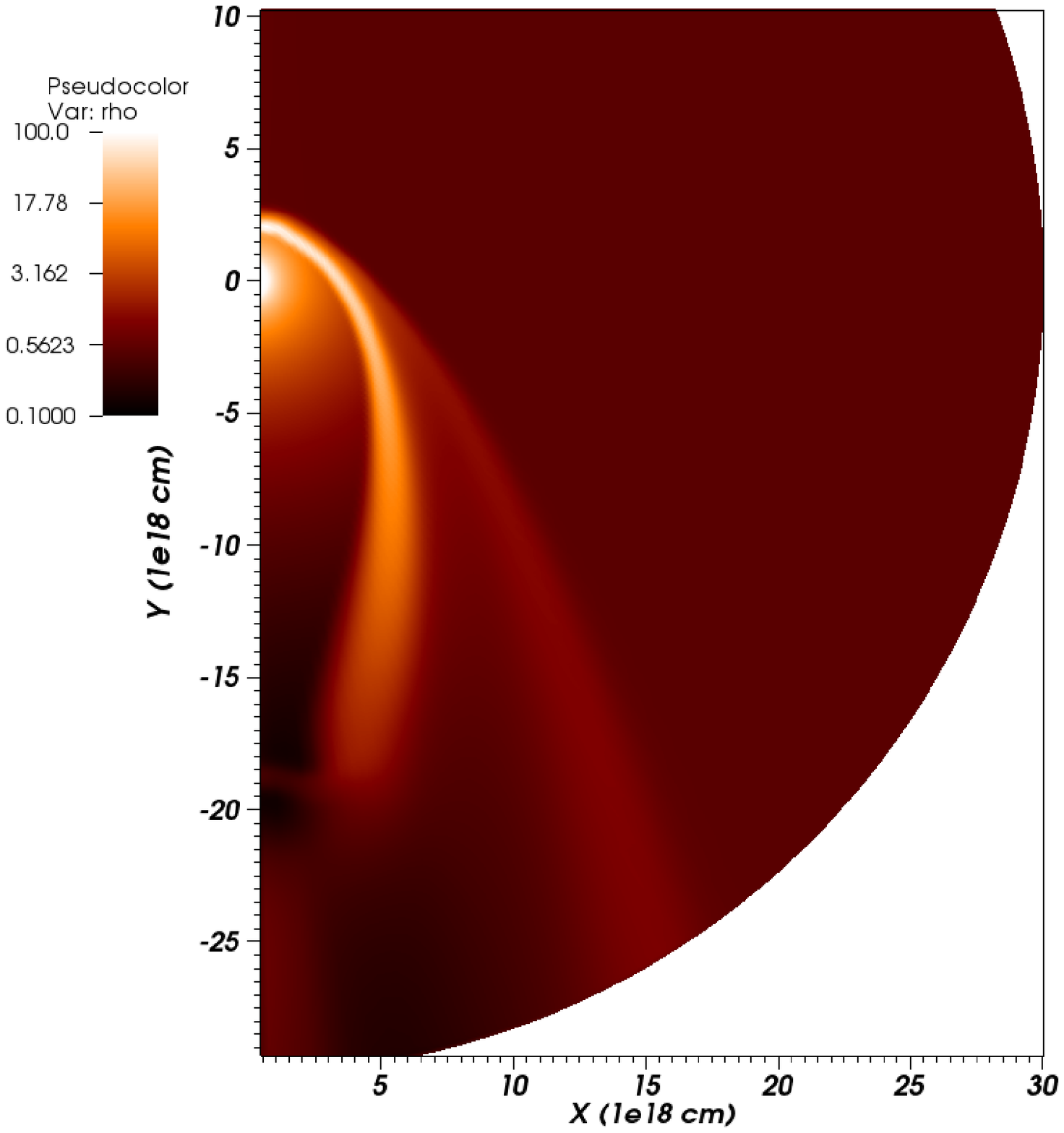} &
\includegraphics[trim=15 50 85 20,clip=true,width=70.0mm,angle=0]{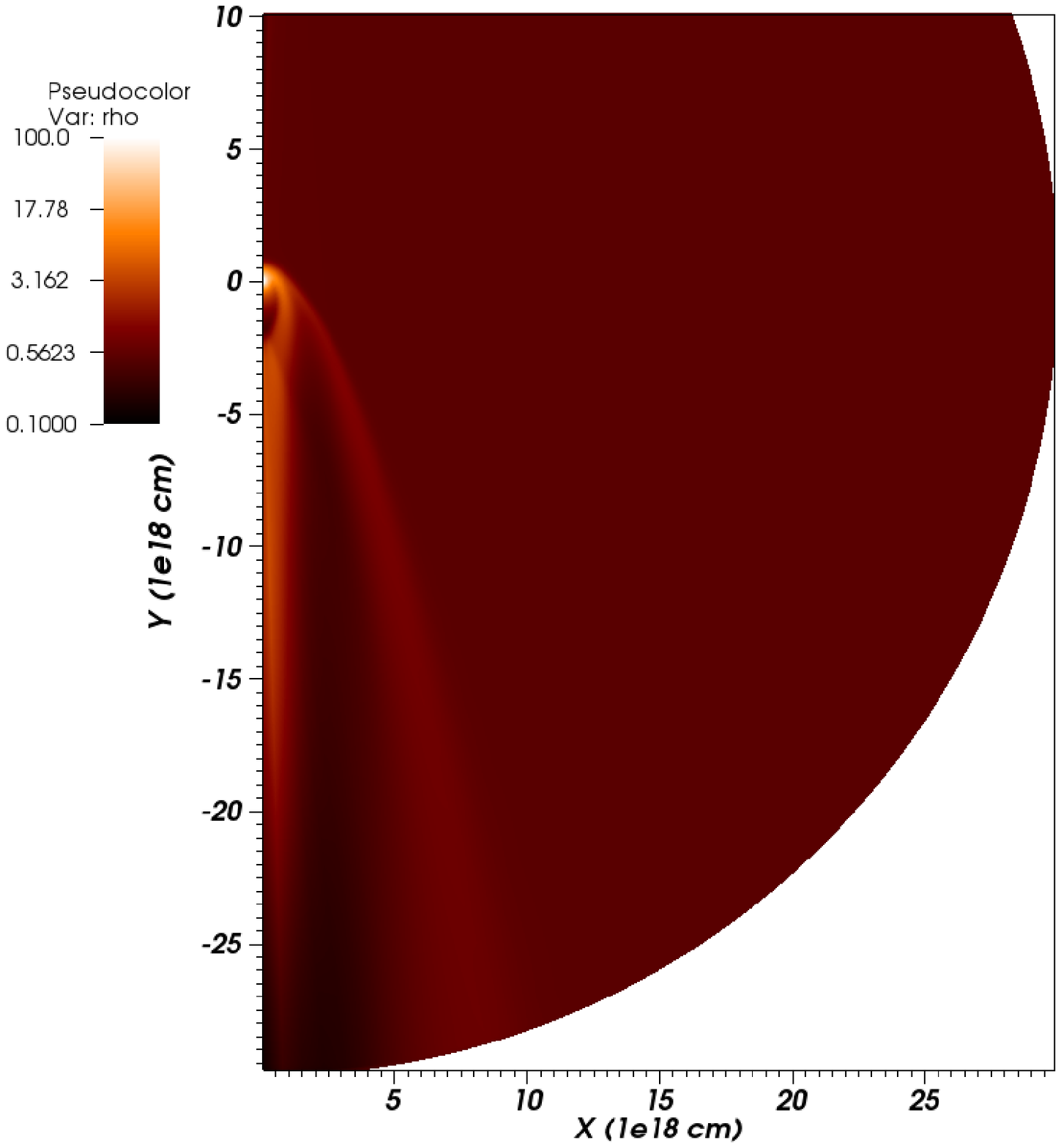} \\
& \includegraphics[trim=15 50 85 20,clip=true,width=70.0mm,angle=0]{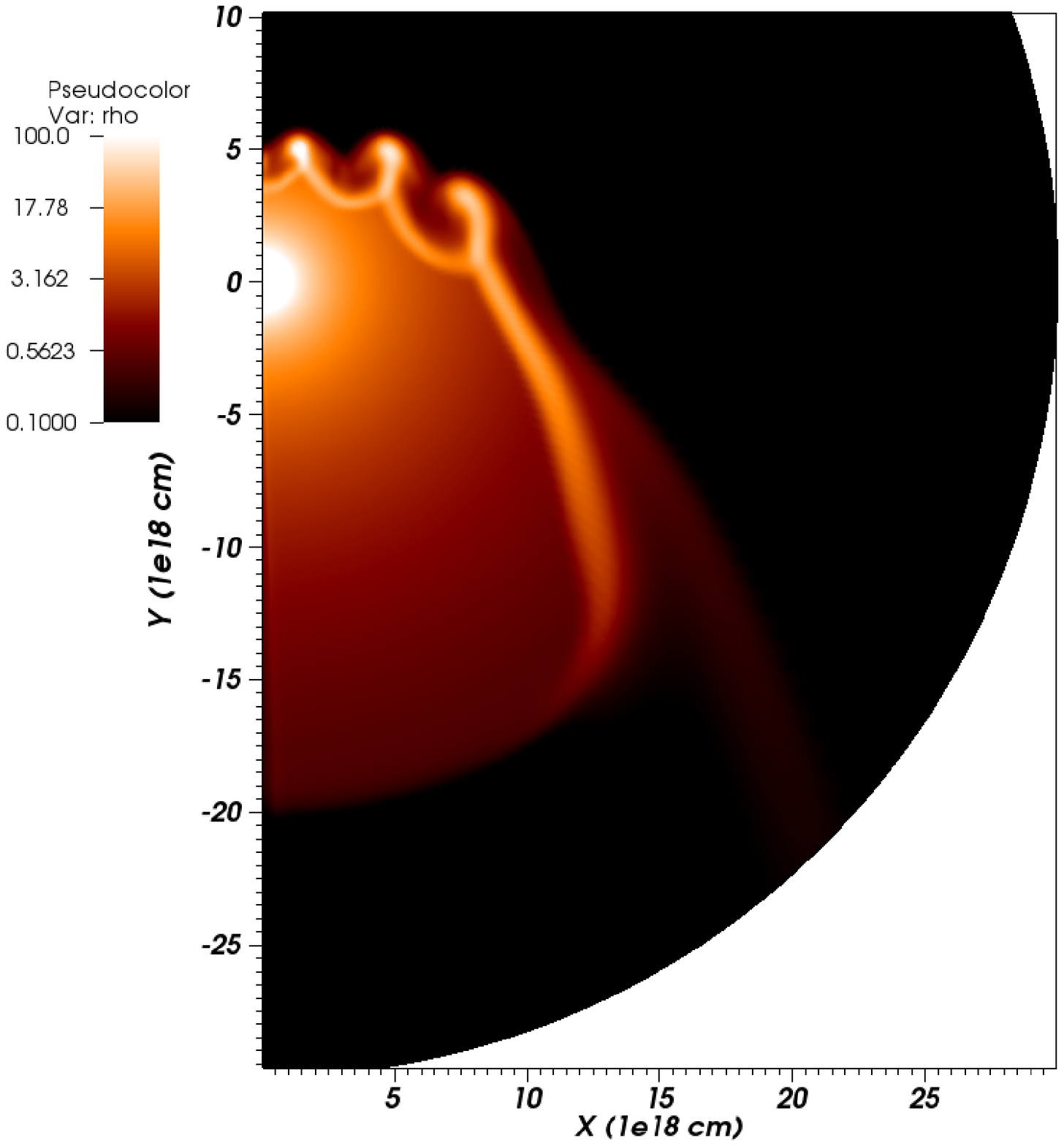} &
\includegraphics[trim=15 50 85 20,clip=true,width=70.0mm,angle=0]{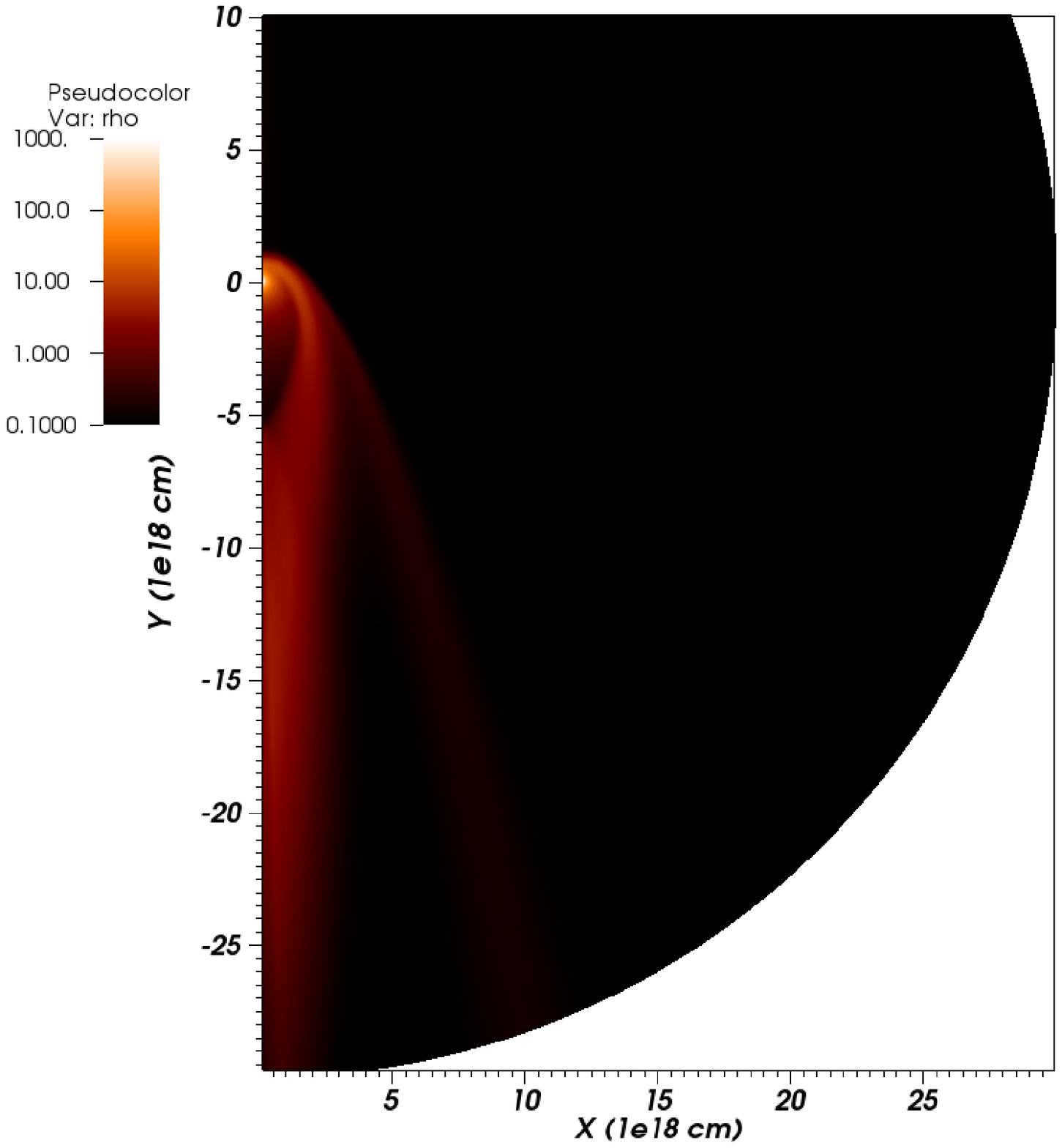} \\
\end{array}$
\end{center}
\caption{The 2D density profile for the four models considering constant wind and AM properties. From left to right, up to down the represented models are: $\rm PN_{stag,Dmin}, PN_{tail,Dmin}, PN_{stag,Dmax}, PN_{tail,Dmax}$ (see Table \ref{tab:Models} and text for details) }
  \label{fig:FailedModels}
\end{figure*}

 Intriguingly, we found that considering time invariant wind and AM properties, none of our models achieved to explain the overall morphology of this PN. The reasons are the following: 
 
 a) For a given possible distance of HFG1, adopting the set of wind/AM properties, which theoretically satisfy the observed position of the stagnation point (see shadowed regions of Fig.~\ref{fig:nMdot}),  the resulting AM ram pressure is not high enough, with respect to the wind flow counterpart,  to form the observed collimated tail which is lying close to the mass-losing object while the resulting overall structure of the outer shell is rather extended and its shape deviates substantianlly from the observed one.  

To illustrate this, we present two models of our grid which correspond to the cases where HFG1 takes the minimum and maximum value among its possible distance range  (Model: $\rm PN_{stag,Dmin}$ for $D_{min}= 310$~pc and Model: $\rm PN_{stag,Dmax}$ for $D_{max}= 950$~pc). The parameters of these two models (see Table \ref{tab:Models}) are chosen to reproduce the observed stagnation point radius of HFG1 for each distance (Eq. \ref{eq:stag}). The resulting density structure of the bow shaped wind bubble for these two models is depicted in the left column of Fig.~\ref{fig:FailedModels}. As predicted by the analytical approximation both models reproduce the observed radius of the stagnation point which is $1.9 \times 10^{18}$~cm = 0.6 pc and $5.5 \times 10^{18}$~cm = 1.8 pc for $D=310$~pc and $D=950$~pc, respectively. Nevertheless, for both models due to the high ram pressure, apart from the radius of the stagnation point, the rest structure of the outer shell is much more extended and its shape is more elongated in the direction of motion than the observed one (see Table \ref{tab:Models_comp}).  Interestingly, Eq. \ref{eq:r/ro} does not satisfy the observed geometry of the PN outer shell suggesting that the wind bubble of HFG1 is out of a steady state (Eq. \ref{eq:tflow}).  As far as the tail properties are concerned, the model $\rm PN_{stag,Dmin}$ results to a wider tail which starts far beyond the CS compared with what we observe, while in the model $\rm PN_{stag,Dmax}$ no tail is formed (Table \ref{tab:Models_comp}).

 b) On the other hand, a collimated flow in the form of a tail close to the CS can be formed either by   increasing the ISM medium ram pressure (increasing the $n_{AM}$) or/and by decreasing this of the wind (decreasing $\dot{M}, u_w$).  The models $\rm PN_{tail,Dmin}$ and $\rm PN_{tail,Dmax}$ (right column of Fig. \ref{fig:FailedModels}) represent this case for the minimum and maximum distance of HFG1, respectively. The width, the length and the starting point of the tail agrees with what we observe at the PN. However, due to the high ram pressure of the ISM, the radius of the stagnation point and the overall wind bubble structure are much smaller than the observed ones (Table \ref{tab:Models_comp}).

\begin{table*}
	\centering
		\caption{The properties of the four studied models with constant wind and AM properties. The $D$ refers to the assumed HFG1 distance of each model, while  $\rm \tau_{AGB}$ refers to the time interval of the wind bubble evolution. At all of the models it has been assumed that the temperatures of the wind and the AM are $T_{wind}= T_{AM}= 10^3$~K. }
	\label{tab:Models}
		\begin{tabular}{l |c|c|c|c|c|c|c }
		\hline
		\hline
		Model &D \rm(pc)& $ \dot{M}~\rm(M_{\odot}~yr^{-1})$& $ u_w~\rm(km~s^{-1})$ & $ u_*~\rm(km~s^{-1}) $ & $ n_{AM}~\rm(cm^{-3})$  & $ \tau_{AGB}~\rm(Myr) $  \\
	\hline
$\rm PN_{stag,Dmin}$	&310& $10^{-5}$ & 5.0 & 30 & 0.5 & 1.1 \\
$\rm PN_{tail,Dmin}$	         &310& $3 \times 10^{-7}$ & 5.0 & 30 & 0.5 & 1.1 \\
$\rm PN_{stag,Dmax}$	&950& $3 \times 10^{-5}$ & 5.0 & 60 & 0.05 & 1.1 \\
$\rm PN_{tail,Dmax}$	&950& $10^{-6}$ & 5.0 & 60 & 0.1 & 1.1 \\

\hline
		\end{tabular}

\end{table*}

\begin{table*}
	\centering
	\caption{Comparison of the observed morphological properties of HFG1 with the relevant extracted values of our modeling for the two distance limits of this object.  $r_{\theta= 90^o}$,   $r_{\theta= 170^o}$ stand for the radius of the outer shell for $90^o$ and  $170^o$ angles between the radius vector and the vector of the stellar velocity while  $w_{tail}$ refers to the width of the tail.   }
		\begin{tabular}{l |c|c|c|c|c|c| }
		\hline
		\hline
		 & $D~\rm(pc)$& $r_0~\rm(10^{18}~cm)$ & $r_{\theta= 90^o}~\rm(10^{18}~cm)$ & $r_{\theta= 170^o}~\rm(10^{18}~cm)$  & $w_{tail} ~\rm(10^{18}~cm) $  \\
	\hline
Observations                     &310&1.8&2.1&2.6&1.6\\
\hline	
$\rm PN_{stag,Dmin}$	&310&1.9&3.5&18.7&5.0\\
$\rm PN_{tail,Dmin}$	         &310&0.5&0.7&2.1&1.6\\
\hline
\hline
Observations                     &950&5.5&6.3&7.9&4.8\\
\hline	
$\rm PN_{stag,Dmax}$	&950&$\sim 5$&8.6&29.4& no tail \\
$\rm PN_{tail,Dmax}$	         &950&0.9&1.8&5.1&4.9\\
\hline
		\end{tabular}
	
	\label{tab:Models_comp}
\end{table*}

 \subsection{Alternative scenaria}\label{sec:Alternatives}

In the previous section we showed that a description of the problem considering time invariant wind and AM properties is not able to explain the overall morphology of HFG1. Precisely, these models cannot reproduce at the same time the  extended bow shaped shell and the collimated tail observed at the PN.

Considering that the overall shape of the bow shaped shell and tail are determined by the pressure balance of  the wind and AM flows, we increase  the complexity of the problem assuming that during the formation of HFG1 the wind or/and the AM ram pressure are time variables. Under this assumption someone could suggest that, during the early phase of the PN evolution  the ratio of the AM ram pressure ($P_{AM}$) over this of the wind ($P_{wind}$) is high enough to compress  the wind material behind the CS into a collimated tail. Subsequently, as time progresses, this ratio decreases -- the wind ram pressure becomes more dominant -- and the bow shaped shell expands to the observed size of HFG1. The PN retains  the tail being a remnant of the previous phase of the PN evolution. 
 Given the variables which are entangled in the description of the wind and ISM ram pressures (see Eq. \ref{eq:ramPress}) such a scenario can occur if during the formation of the PN at least one of the following processes are taking place:   $i)$~the CS systemic velocity decreased,  $ii)$~the AM density decreased, $iii)$~the wind mass loss rate or/and its terminal velocity increased. In this section we test the possibility and the plausibility of such a process to occur during the HFG1 evolution. 

{\bf Case (i):} A rapid change at the systemic velocity of a star or binary system is possible either through a collision with an other star or binary \citep{Portegies99} or by being a member of a binary or triple system and receiving a kick velocity due to a rapid mass loss episode of its companion star \citep[e.g. a supernova explosion,][]{Blaauw61}.   Thus, case (i)  implies that during the formation of HFG1, V664 Cas either experienced a collision with an other star/binary or was a triple system in which the third member exploded as a supernova. Such an interaction should occur in a way to decelerate the object leading to the expansion of the outer wind shell.  Nevertheless, the specific conditions that this scenario demands, which should take place in the short time duration of the PN formation ($t_{PN} \sim 1$~Myr), combined with the lack of any observational evidence that any of the aforementioned processes occurred (e.g. traces of a supernova remnant) render case (i) rather unlikely and implausible. 

{\bf Case (ii):} This option suggests  that the AM around HFG1  is characterized by a sharp negative density gradient in the direction of HFG1 motion. Thus, during the formation of the PN, HFG1 has passed from a medium with a specific density to an other with a lower one. The decrease of the AM ram pressure due to the decrease of its density leads to the expansion of the outer shell at the final stages of the PN evolution.  Such a scenario has been suggested in  literature to explain the formation of the wind bubble that surrounds the Mira AB binary system which reveals a structure similar to HFG1 described by  an extended bow shock  accompanied by a collimated tail \citep{Esquivel10, Wareing12}. 

Nevertheless, this scenario reveals the following weakness: the pressure balance at interface of the high and low density media that this scenario requires, demands the temperature of the  low density medium to be higher than this of the former.  In other words, case (ii) suggests that HFG1 has passed from a warm neutral medium into a hot, low density medium.  Due the high temperature, the sound speed in the hot medium is also high (e.g. for $n=10^{-3}~cm^{-3}$ and $T=10^6$~K we get a sound speed of $c_s \sim 70~km~s^{-1}$). This high sound speed results in a subsonic/transonic velocity shear at the wind/AM interface. It has been shown that the spreading rate of transonic/subsonic mixing layers is much larger than this  of high Mach number mixing layers \citep{Canto91}. This results in a complex, turbulent structure of the bow shock/tail bubble \citep{Wareing07, Esquivel10}, something that is not observed in HFG1. 

{\bf Case (iii):} The final option suggests that the wind ram pressure follows the increase of the mass loss rate or/and the wind terminal velocity.  As we explained in the Introduction,  PNe are formed by  mass outflows that accompany the ending phase of AGB stars.  At this evolutionary stage the stars are characterized by strong {\it time variable winds} with mass loss rates which increase by at least a factor of $10^2 - 10^3$ from the early AGB phase up to the final thermal pulsating and super wind phase \citep{Vassiliadis93}. Thus, case (iii) offers a natural explanation of a progressive wind ram pressure domination aligned with the stellar evolution theory and, in contrast to cases (i) and (ii), it does not demand specific  external  conditions/processes that coincidentally occurred during the PN formation.  Thus,  this scenario seems to be the most plausible and worths further investigation.

\subsection{Models with time depended wind properties}

\subsubsection{Estimating the PN progenitor properties based on the observables of V664 Cas primary star}\

\begin{figure}
\includegraphics[trim=0 0 0 0,clip=true,width=\columnwidth,angle=0]{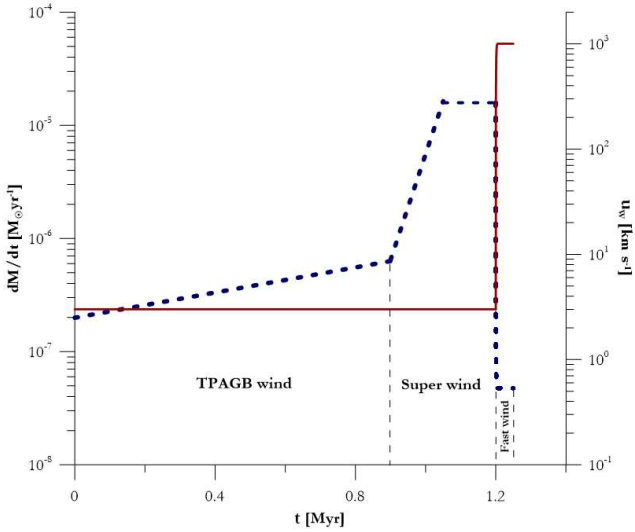} 
\caption { The time evolution of the wind mass loss rate (black dashed line) and terminal velocity (red solid line) over the formation of the planetary nebula as described by Eq. \ref{eq:dotM} (AGB wind) and Eq. \ref{eq:dotM_fast} (fast wind).}
\label{fig:dotM}
 \end{figure}

 In the previous section we argued that the cometary structure of HFG1, could possibly be reproduced by a time variable wind of its progenitor star. In order to keep our model self-consistent, based on the observed properties of its central star, V664~Cas, we first attempt to estimate the initial stellar mass and metalicity of the PN progenitor. Then we describe its wind properties and variability during the formation of the PN by adopting the predictions of the stellar evolution theory.

The primary of V664~Cas, which formed the PN, is a sdO  star. Its high estimated effective temperature, $T_{eff} = (8.3 \pm 0.6)\times 10^4$~K, in combination with its estimated surface gravity  $\log g = 5.65 \pm 0.05~\rm cm~s^{-2} $ (see Table \ref{tab:V664Cas_prop}) classifies this object as a {\it luminous} sdO \citep{Napiwotzki08}.  This subclass of sdO  stars is consistent with a post-AGB or post-early-AGB nature \citep{Schoenberner83,Bloecker95,Napiwotzki08}.

The initial mass of the sdO progenitor  star can be estimated through the comparison of the current, observed properties of V664~Cas primary star with the evolutionary tracks of AGB/post-AGB stars by \citet{Schoenberner83} and \citet{Bloecker95}. According to \citet{Shimanskii04} the mass of the sdO primary star is in the range of 0.54 - 0.60 \msun.  Within this mass range the post-AGB tracks indicate a 1 - 3 \msun~AGB progenitor, respectively. The comparison between the same evolutionary tracks and the observed properties of the sdO  star   in the surface gravity - effective temperature diagram \citep{ Napiwotzki08}  favors more the lower band of the aforementioned initial mass range. However, given the uncertainties that include both the observational derivation of the V664~Cas fundamental parameters and the estimated post-AGB tracks we adopted a broader range of initial masses. In particular we  test the cases of an AGB wind progenitor with initial masses 1, 2, 3 and 4 \msun  \footnote{We did not extent the chosen mass range to lower masses as for stars with initial masses $\lesssim$ 1\msun~the ejected material during the AGB phase is small and the post-AGB evolution is very slow something that prevents the formation of a PN \citep{Napiwotzki08,Otsuka15}.}. Finally, the initial metallicity of the sdO progenitor is considered to be solar (Z = 0.02) since the F5-K0 main sequence secondary star of V664~Cas reveals roughly solar composition \citep{Exter05}.

Here,  we present the model that best reproduces the properties of HFG1 and this corresponds to an AGB star with initial mass of 3 \msun. However, note that the cases of 2~\msun~and 4~\msun~revealed similar results and also  -for the proper set of parameters- can roughly reproduce the observable of HFG1. This is not happening for the case of a 1~\msun~AGB wind as for such a  star the AGB evolution terminates immediately before reaching the thermal pulsating AGB phase \citep{Bloecker95}. As a result the rapid increase towards high mass loss rates is absent in this case and hence, is not possible to form the extended bow shaped shell observed in HFG1.

\begin{table*}
	\centering
		\caption{The properties of the three studied models  based on the wind properties of a 3 \msun~AGB star. At all of the models it has been assumed that the temperatures of the wind and the AM are $T_{wind}= T_{AM}= 10^3$~K. }

		\begin{tabular}{l |c|c|c|c|c|c|c }
		\hline
		\hline
		Model &$D$ \rm(pc)& $ \dot{M}~\rm(M_{\odot}~yr^{-1})$& $ u_w~\rm(km~s^{-1})$ & $ u_*~\rm(km~s^{-1}) $ & $ n_{AM}~\rm(cm^{-3})$  & $ \tau_{AGB}~\rm(Myr) $  \\
	\hline
$\rm PN_{3M_{\odot},Dmin}$  &310& see Eq. \ref{eq:dotM} & 3.0 & 30 & 0.5 & 1.2  \\
$\rm PN_{3M_{\odot},Dopt}$  &490& see Eq. \ref{eq:dotM} & 5.0 & 35 & 0.2 & 1.2  \\
$\rm PN_{3M_{\odot},Dmax}$  &950& see Eq. \ref{eq:dotM} & 12.0 & 50 & 0.04 & 1.2\\  

\hline
		\end{tabular}
	\label{tab:Modelswindvar}
\end{table*}

\subsubsection{The formation of the PN from a 3 \msun~AGB time variable wind }

To describe the time variable wind mass loss rates of such a star $(M,Z) = (3~\rm M_{\odot}, 0.02)$ during the thermal pulsating AGB (TP-AGB) phase we use the results of the TP-AGB evolutionary models  computed with the  code {\sc colibri} \citep{Marigo13} and presented at \citet{Nanni13} (see their fig. 3).   By doing this we neglect the effects that arise from the duplicity of the progenitor system as the overall morphology   of HFG1 does not reveal any particular structure or peculiarity (e.g.  bipolar structure or/and strong equatorial outflows) which  demands the consideration of strong binary interactions.  For computational simplicity also we neglect, the short time variability of the mass loss rate over the thermal pulse cycle (the depicted spikes that result form the He-shell flash ignition/quiescence) as they do not greatly affect the final outcome of the simulation \citep{Wareing07}. Then, we interpolate  during the nuclear evolution of the wind mass loss rate using three exponential functions  ($\dot{M} \propto 10^{t}$) which describe the TP-AGB  and the super wind phase (see also Fig. \ref{fig:dotM}):

\begin{equation}\label{eq:dotM}
\frac{\dot{M}}{(\rm M_{\odot}/yr )}=  \left\{  
\begin{array}{cc} 
10^{-6.7} \times 10^{~t/1.80} &  \text{for $t= 0 - 0.9$~Myr},\\
\\
10^{-14.7} \times 10^{~t/0.107} &  \text{for $t= 0.9 - 1.05$~Myr},\\
\\
10^{-4.9} &  \text{for $t= 1.05 - 1.16$~Myr}.
\end{array}
\right.
\end{equation}
For simplicity we consider the wind terminal velocity constant as it does not change significantly during the AGB evolution \citep{Vassiliadis93}.\\

 \begin{figure*}
\begin{center}$
\begin{array}{ccccc}
\bf ~~~~~~D = 310~ \rm pc :\\
\\
\bf n ~\rm(cm^{-3})\\
\includegraphics[trim=5 210 513 20,clip=true,width=12.0mm,angle=0]{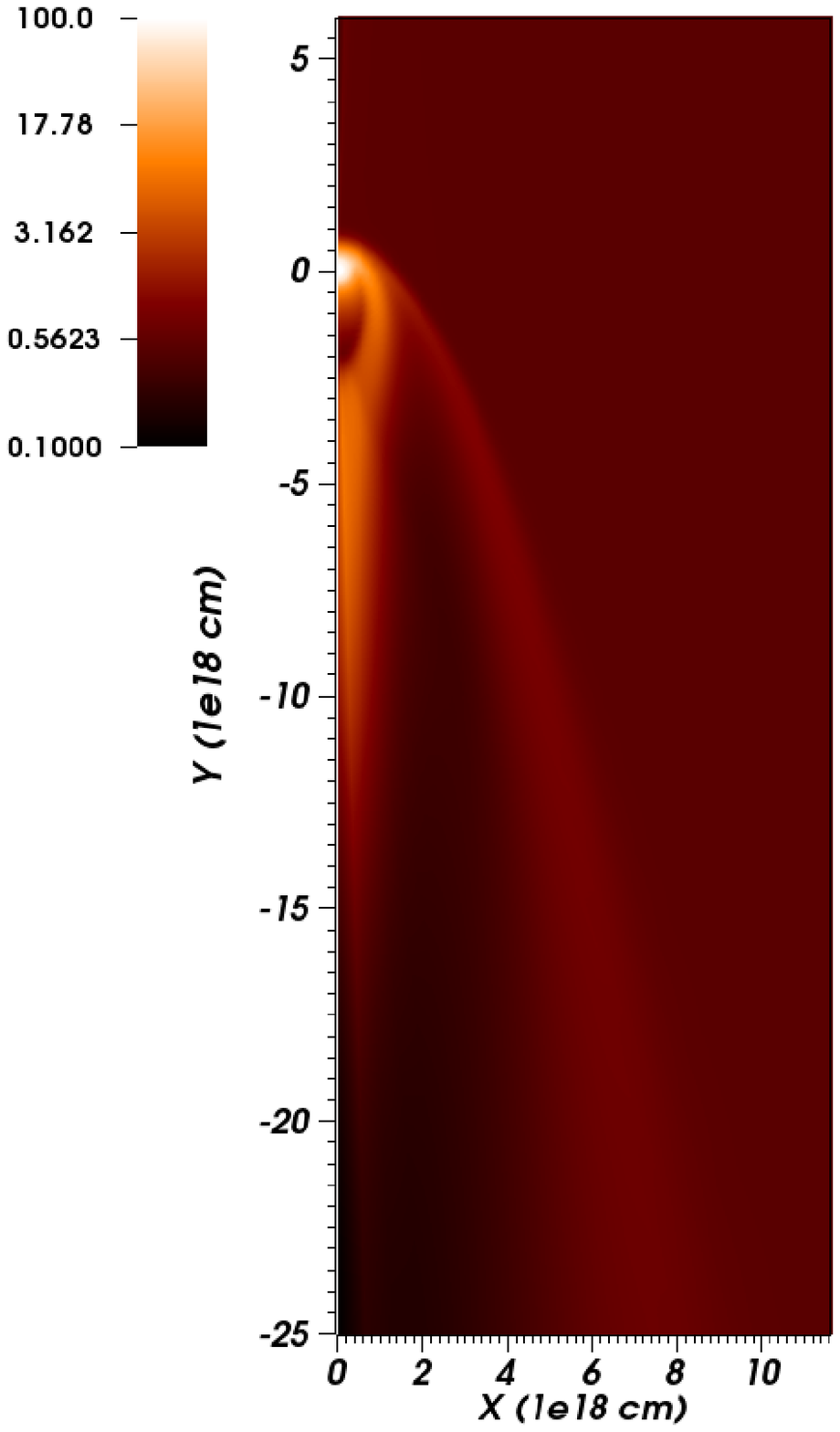} &
\includegraphics[trim=65 50 265 20,clip=true,width=33.0mm,angle=0]{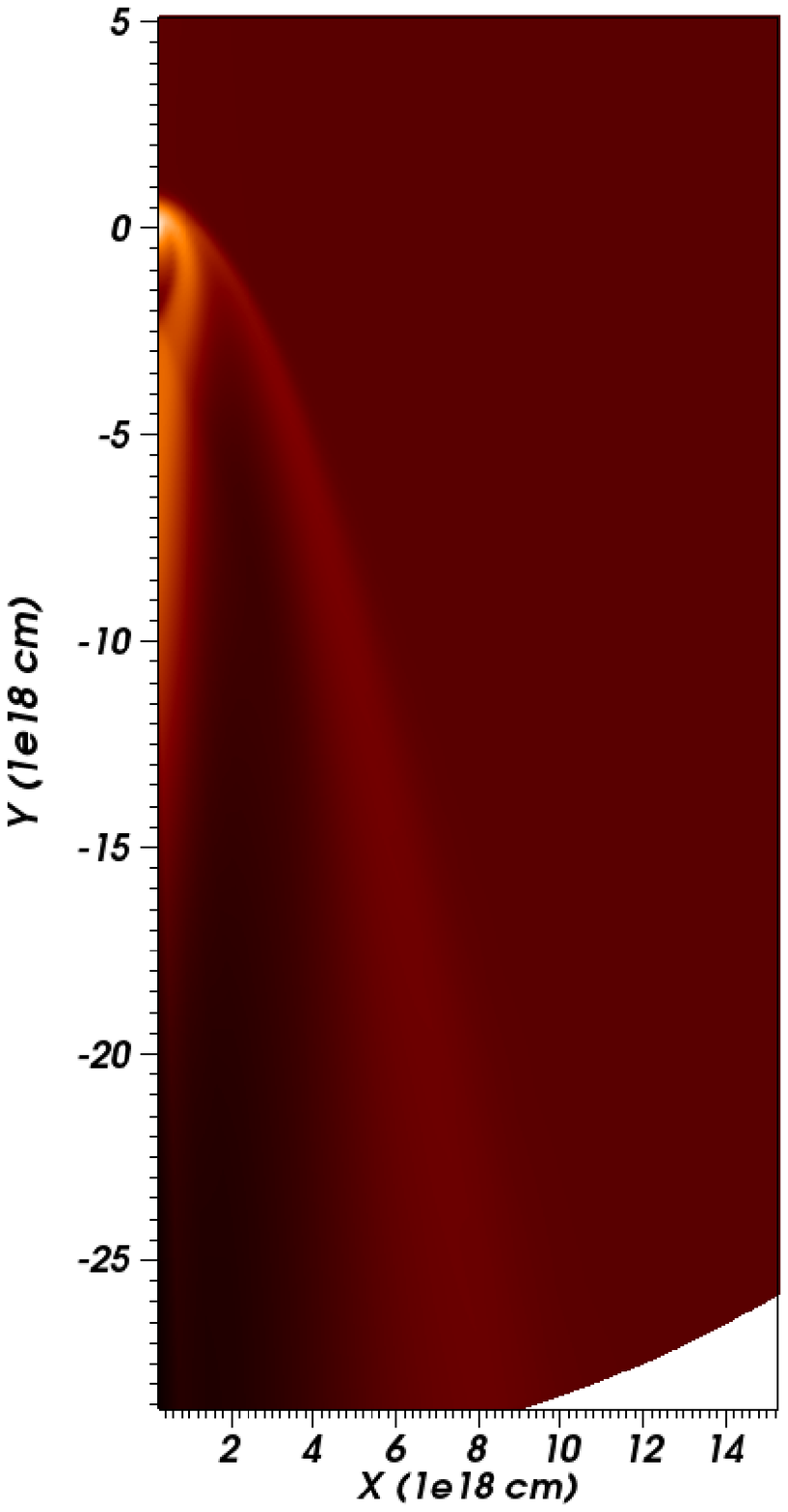} &
\includegraphics[trim=65 50 265 20,clip=true,width=33.0mm,angle=0]{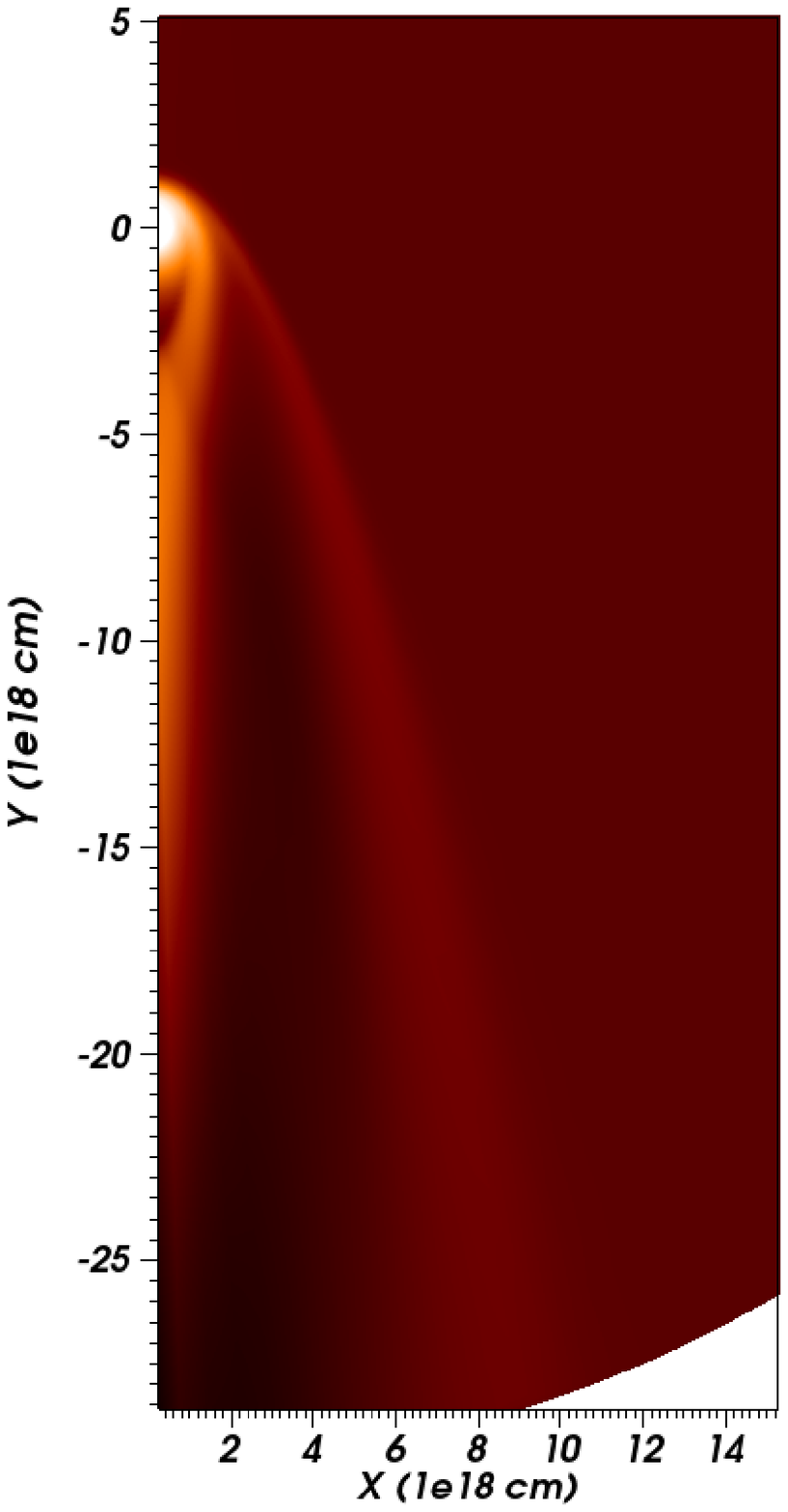} &
\includegraphics[trim=65 50 265 20,clip=true,width=33.0mm,angle=0]{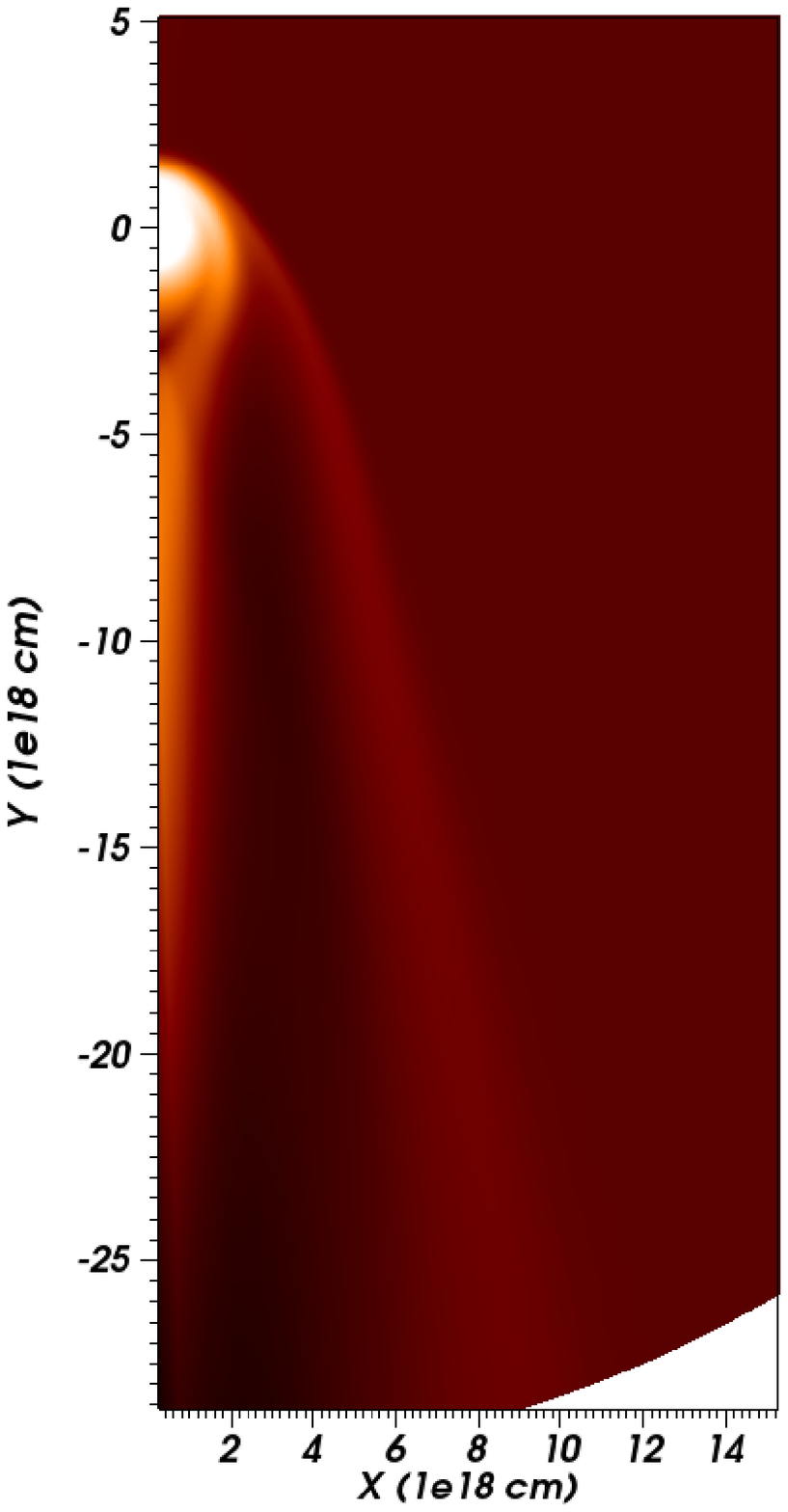} &
\includegraphics[trim=65 50 265 20,clip=true,width=33.0mm,angle=0]{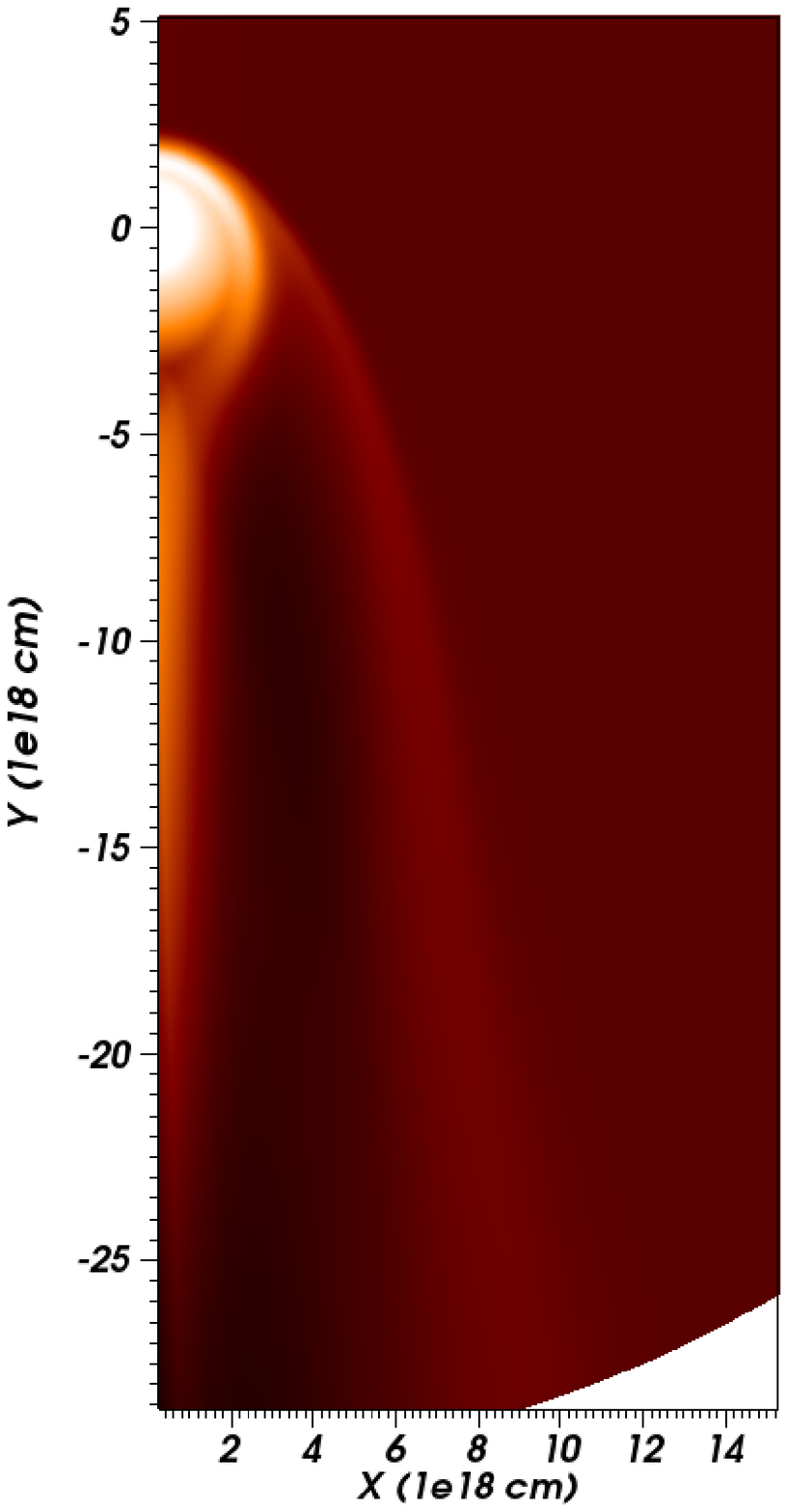}\\
\bf ~~~~~~D = 490~ \rm pc :\\
\\
\bf n ~\rm(cm^{-3})\\
\includegraphics[trim=5 210 513 20,clip=true,width=12.0mm,angle=0]{images/time_evol_20} &
\includegraphics[trim=65 50 265 20,clip=true,width=33.0mm,angle=0]{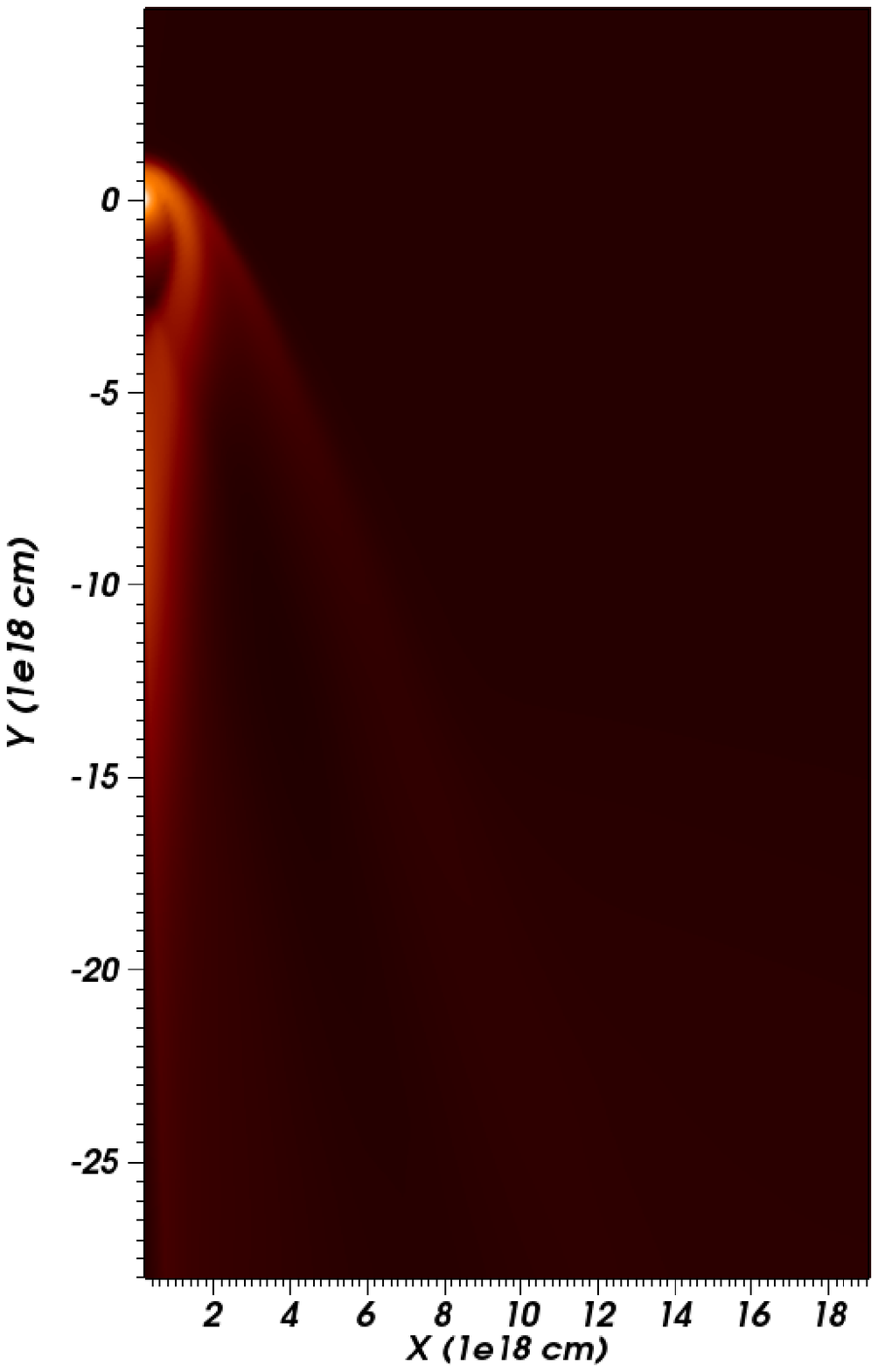} &
\includegraphics[trim=65 50 265 20,clip=true,width=33.0mm,angle=0]{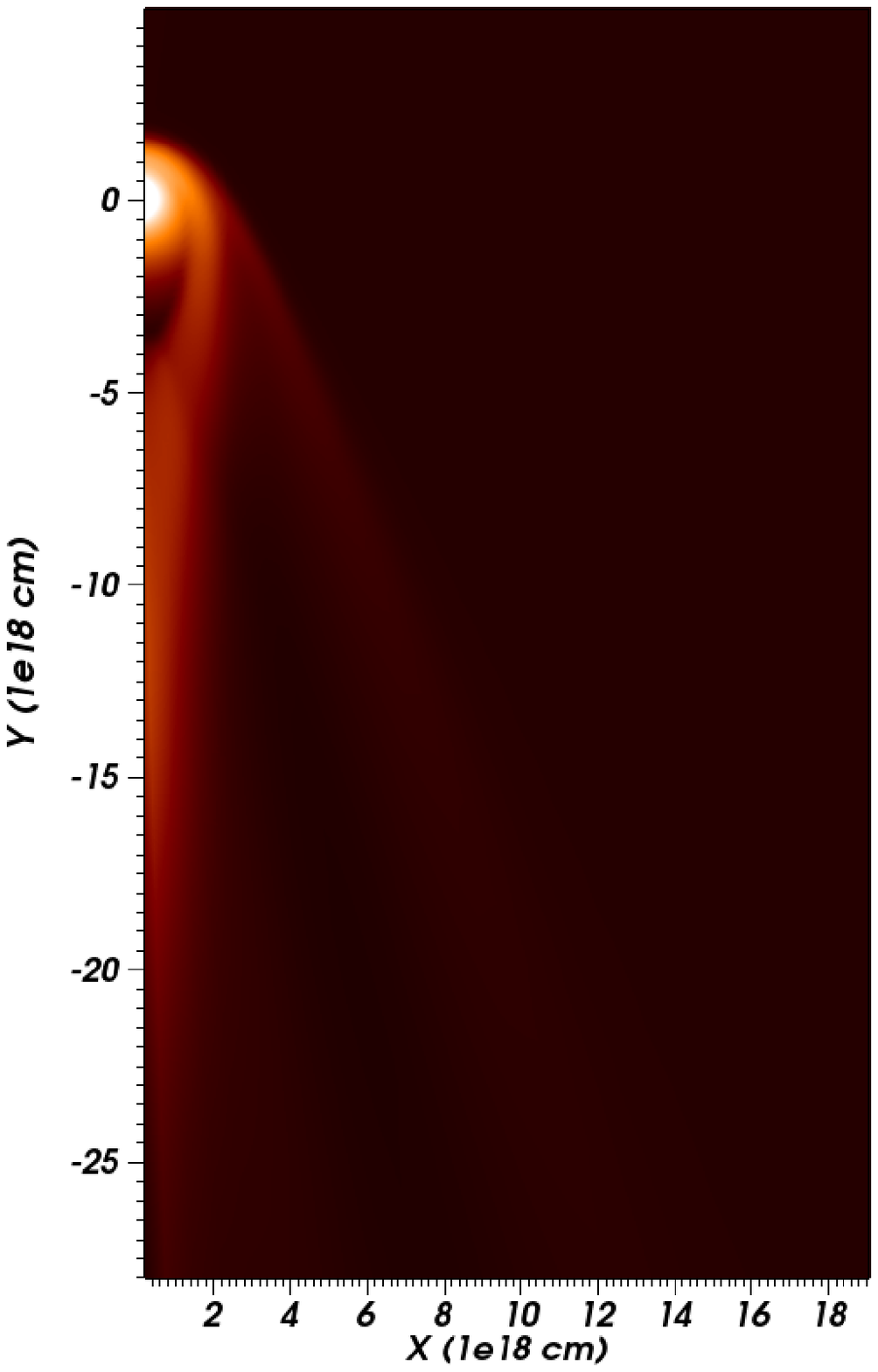} &
\includegraphics[trim=65 50 265 20,clip=true,width=33.0mm,angle=0]{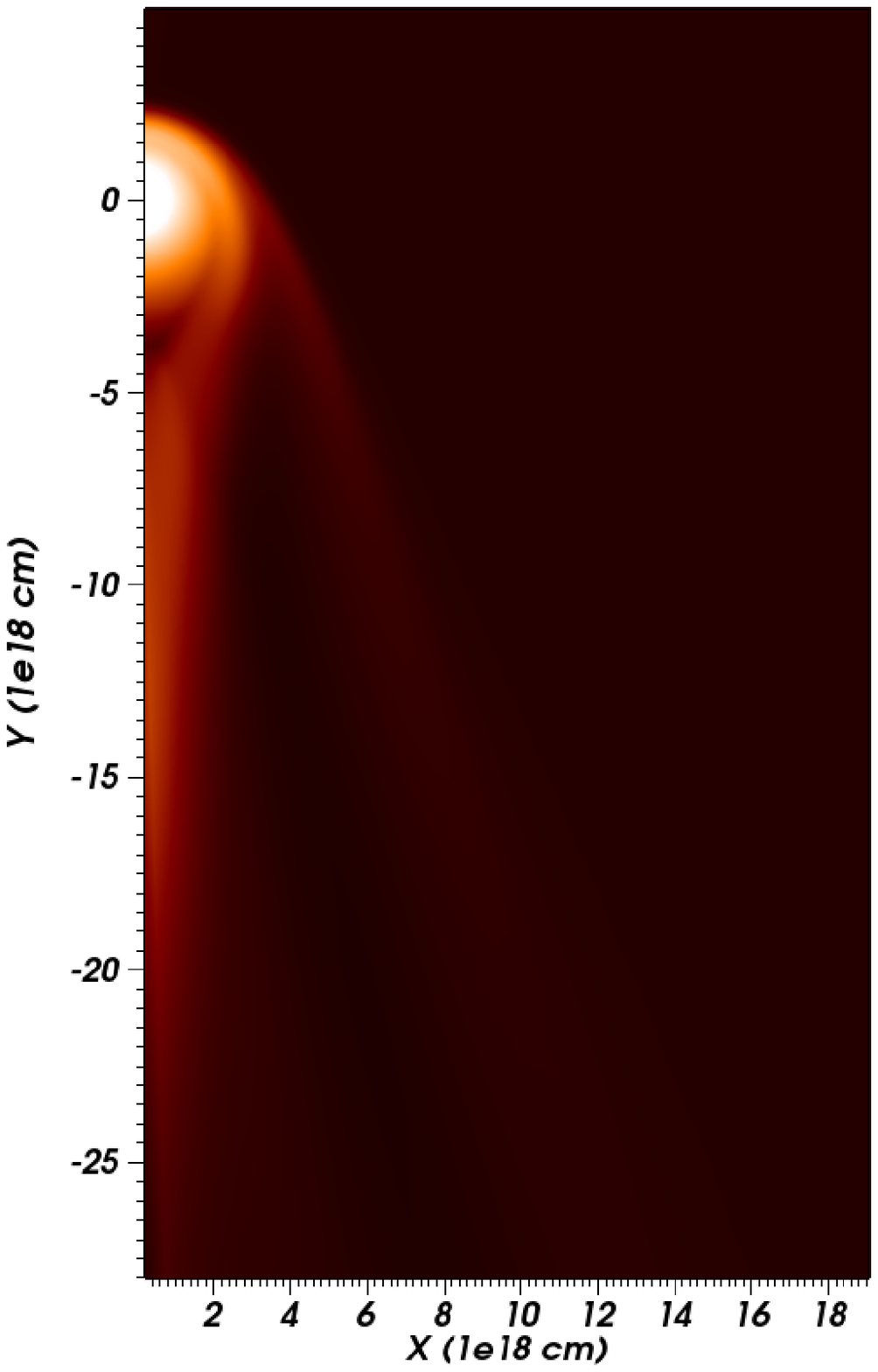} &
\includegraphics[trim=65 50 265 20,clip=true,width=33.0mm,angle=0]{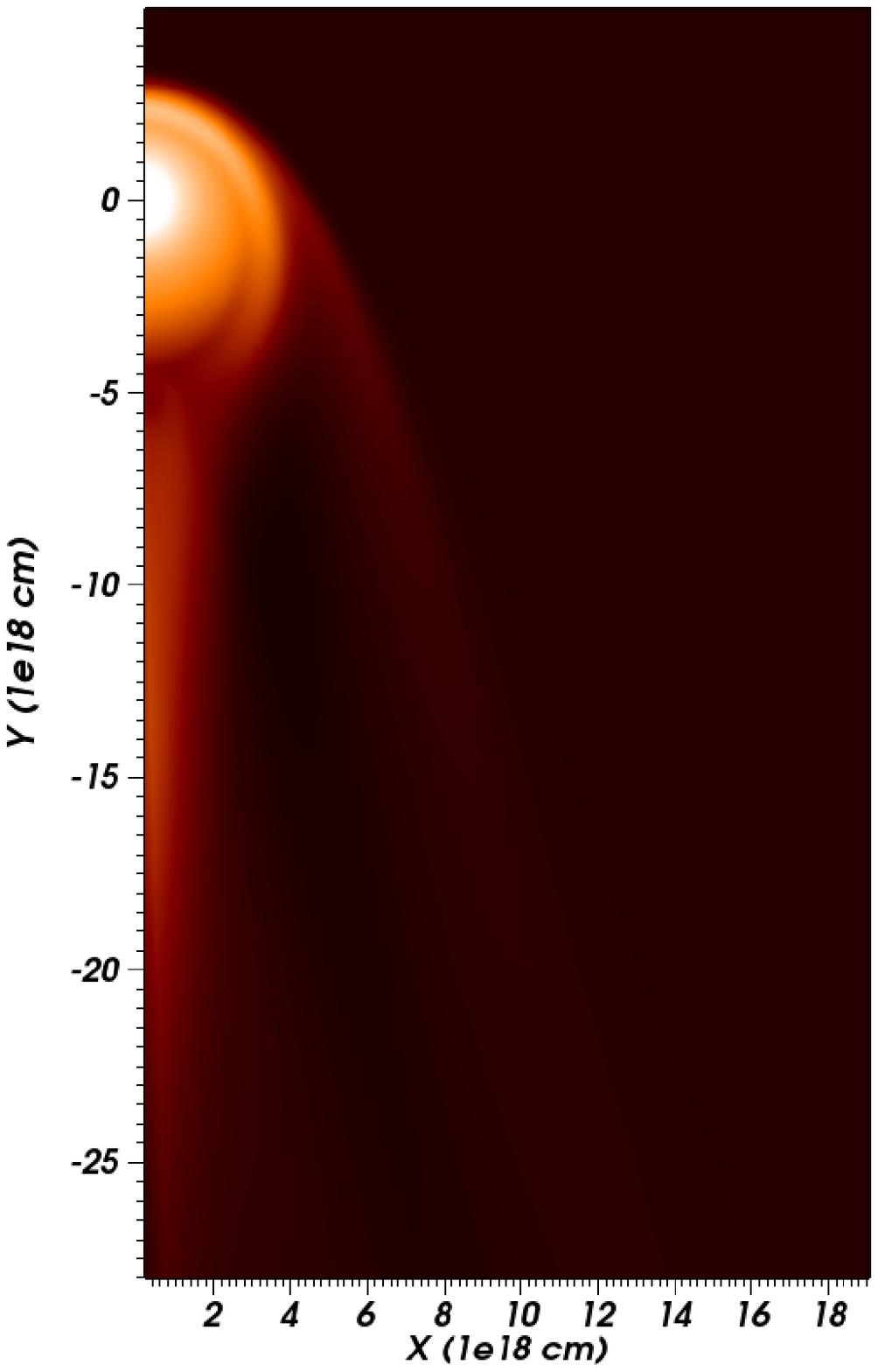}\\
\\
\bf ~~~~~~D = 950~ \rm pc :\\
\\
\bf n ~\rm(cm^{-3})\\
 \includegraphics[trim=5 180 513 40,clip=true,width=12.0mm,angle=0]{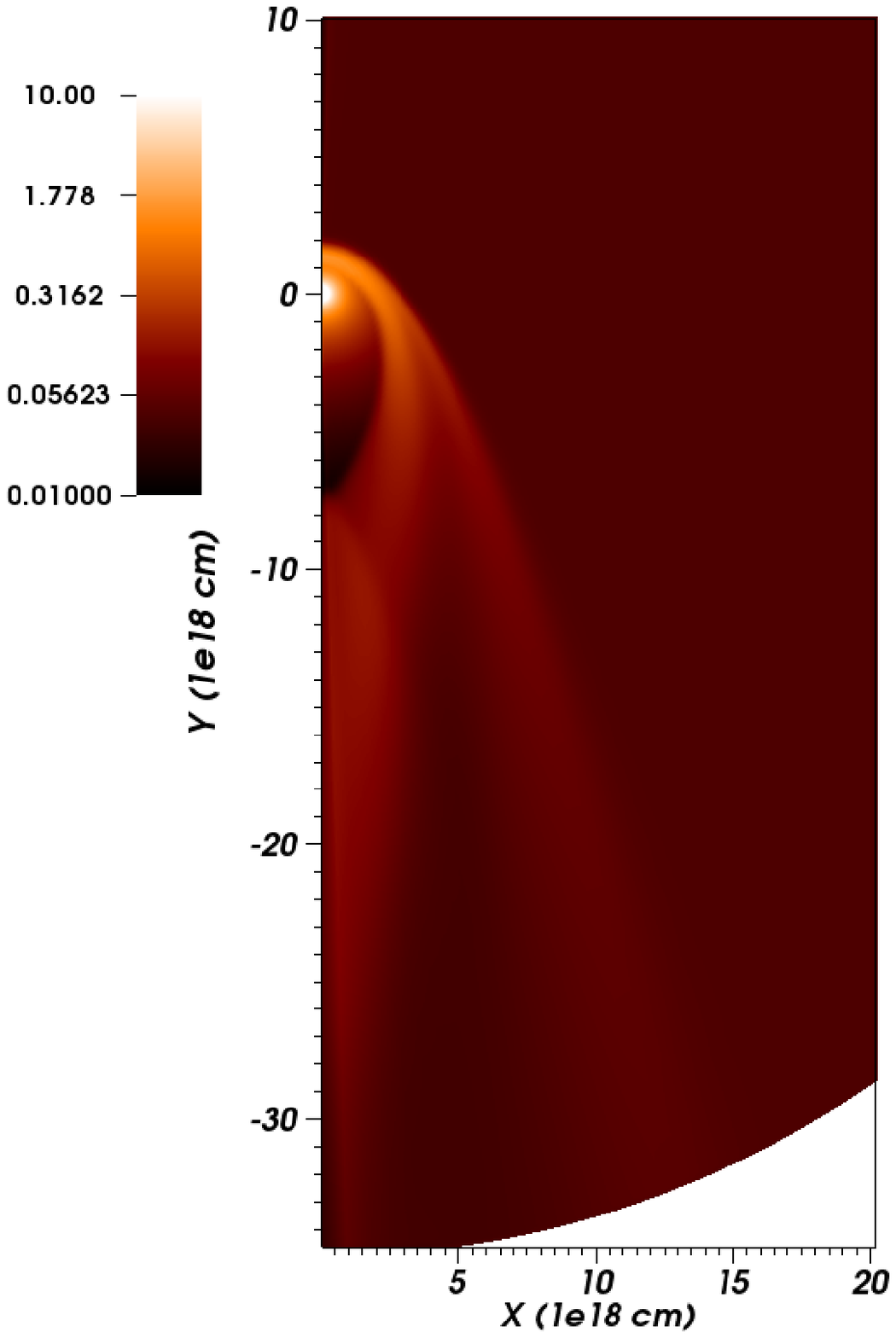} &
\includegraphics[trim=65 50 265 20,clip=true,width=33.0mm,angle=0]{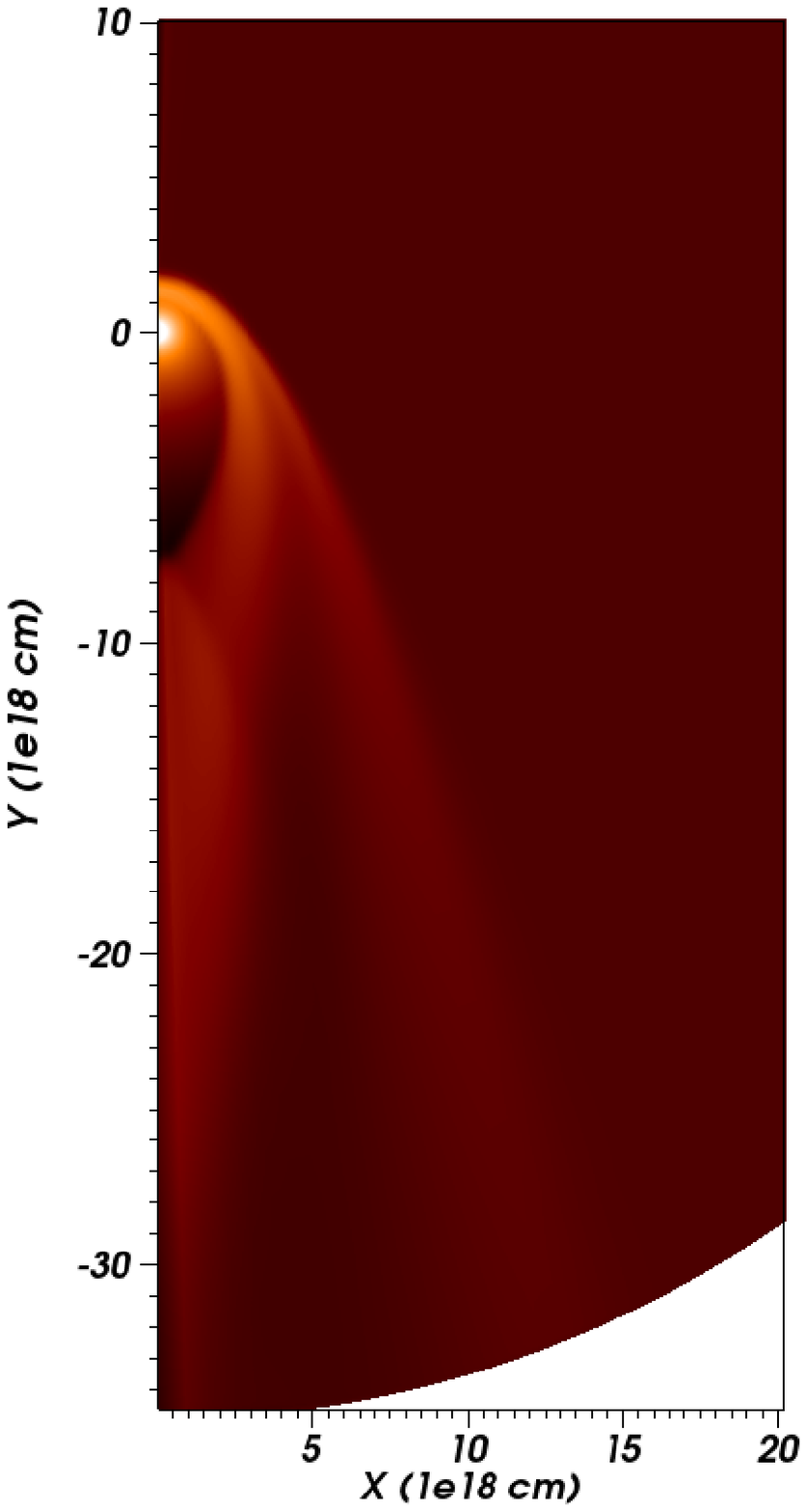} &
\includegraphics[trim=65 50 265 20,clip=true,width=33.0mm,angle=0]{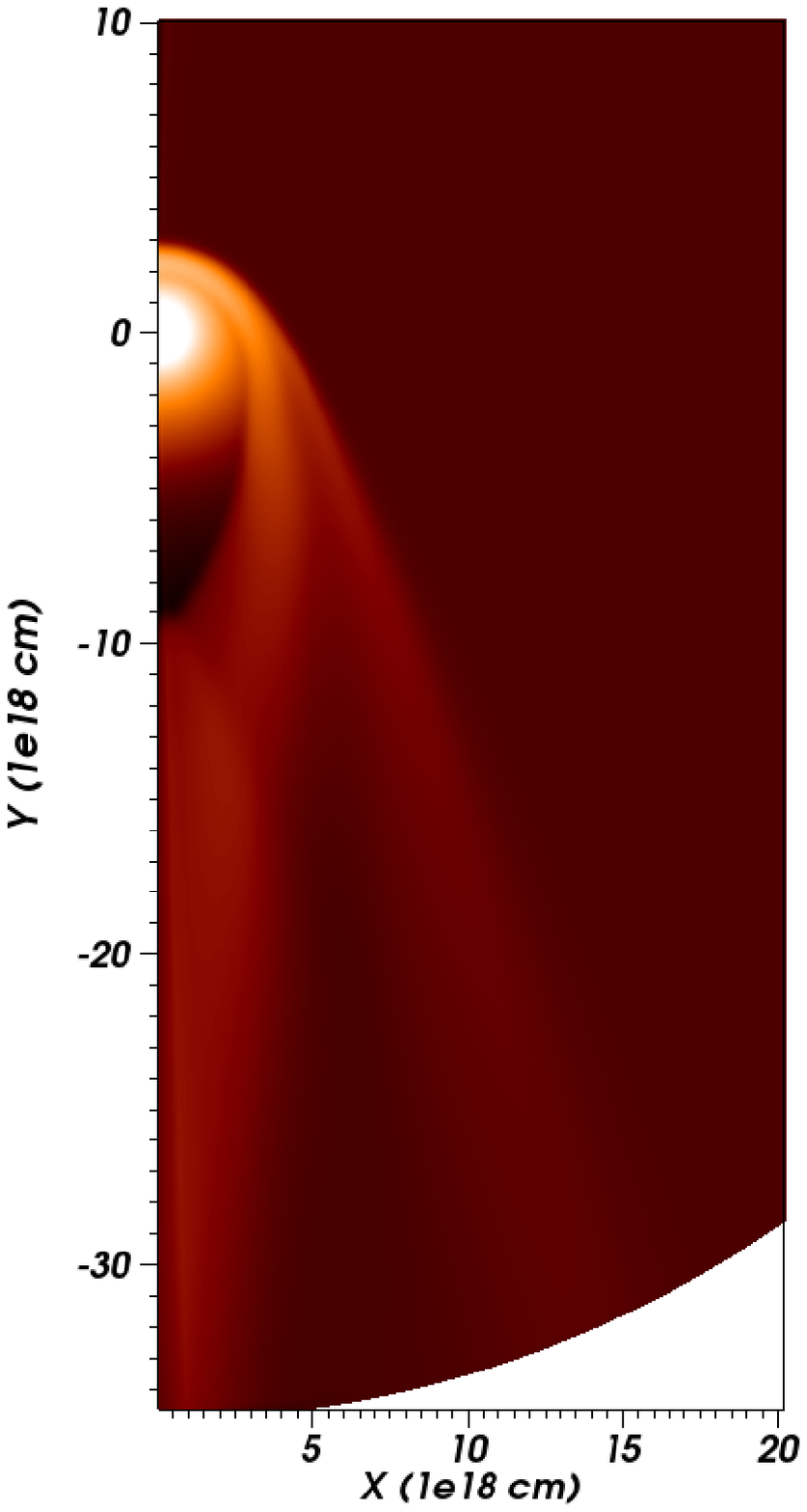} &
\includegraphics[trim=65 50 265 20,clip=true,width=33.0mm,angle=0]{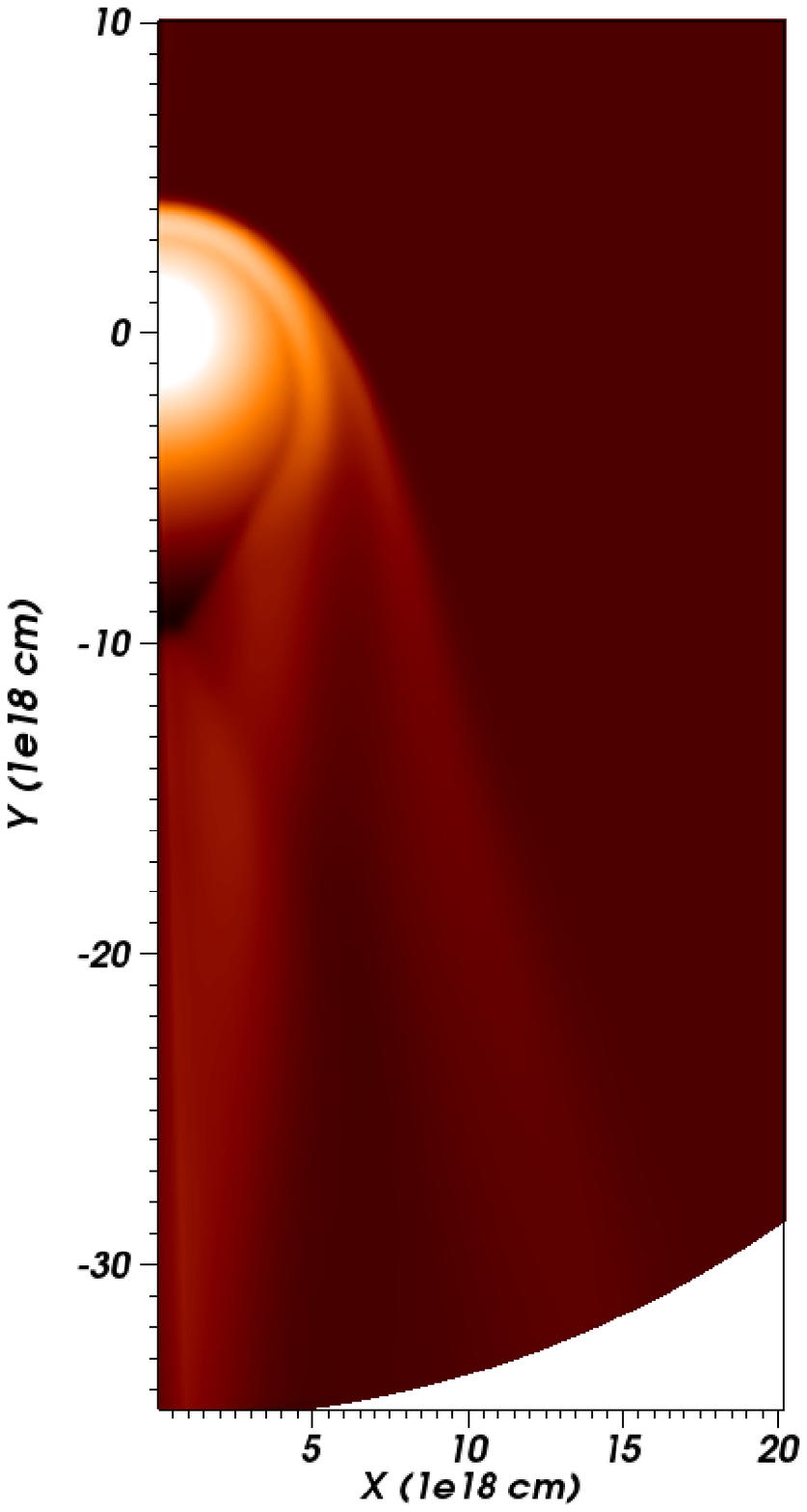} &
\includegraphics[trim=65 50 265 20,clip=true,width=33.0mm,angle=0]{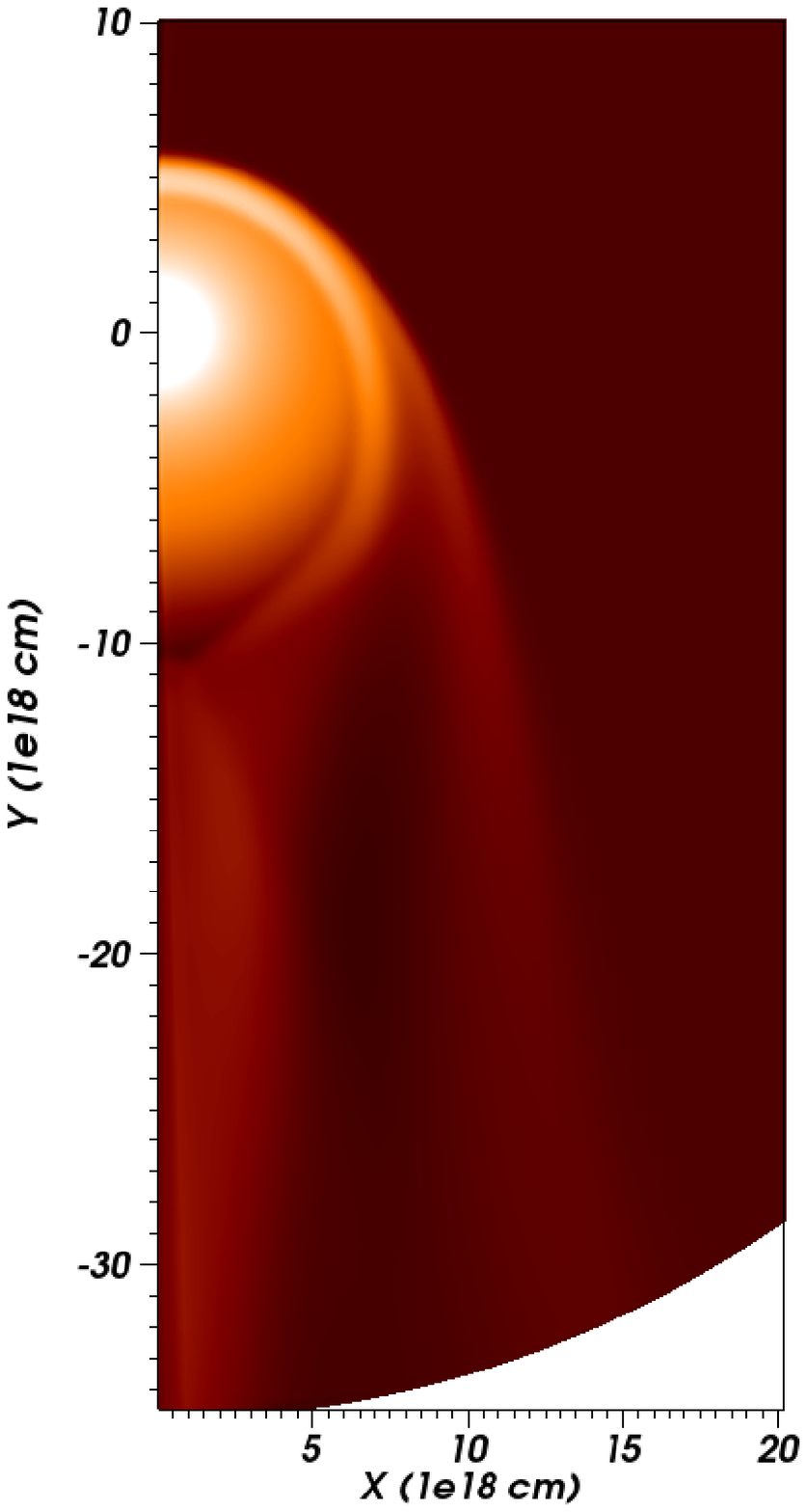}\\
\end{array}$
\end{center}
\caption{The evolution of the bow shaped wind bubble of models $\rm PN_{3M_{\odot},Dmin}$  (upper row), $\rm PN_{3M_{\odot},Dopt}$  (middle row)  and $\rm PN_{3M_{\odot},Dmax}$ (lower row). The snapshots from left to right correspond to the times 0.62 Myr, 1.03 Myr, 1.12 Myr, 1.16 Myr.   }
  \label{fig:bubbleEvol}
\end{figure*}

 Defining the mass loss rate from the aforementioned stellar evolution model, the free parameters of the hydrodynamic modeling are the AM density and the wind terminal velocity  where for AGB stars the later is bounded within the small range of $\sim 5 - 15~ \rm km~s^{-1}$. As we discussed in the previous section the geometrical size as well as the systemic velocity of HFG1 are distant dependent. Thus, any hydrodynamical model should refer to a specific distance of the object. Here we study three cases where at the first two we consider that HFG1 is placed at the minimum and maximum possible distance as estimated by \citet{Exter05}: models $\rm PN_{3M_{\odot},Dmin}$ and $\rm PN_{3M_{\odot},Dmax}$  for $D= 310$~pc and  $D= 950$~pc, respectively. For the third case we consider the PN to be at the optimum distance  estimation as calculated in this work (see Sect. \ref{sec:distance}): model $\rm PN_{3M_{\odot},D_{opt}}$ for $D= 490$~pc.  The aim of considering in our modeling these three distances is to place possible limitations on the distance of the object through the comparison of the models output with the current observables of HFG1. The input parameters of the models which best describe the observed properties of HFG1  are presented in Table \ref{tab:Modelswindvar}.

  Fig. \ref{fig:bubbleEvol} shows the time evolution of the bow shaped wind bubble of these three models. Initially ($t=$ 0.62~Myr) due to the  low wind mass loss rates, the ratio of AM/wind ram pressure is high resulting in a narrow bow shaped bubble followed by a collimated tail. As time progresses ($t=$ 1.03, 1.12, 1.16 Myr) the wind mass loss rate, and hence its ram pressure, increases following Eq. \ref{eq:dotM}. The termination shock expands rapidly forming an extended bow shock. Nevertheless, the changes of the  mass loss rate that occur in the wind source are not communicated instantaneously to the bow shock but need a time interval higher than $t_{flow}$ (see Eq. \ref{eq:tflow}). As the radius of the bow shock increases with the azimuthal angle ($\theta$),  any change at the wind properties is first communicated at the region close to the stagnation point, and then to larger azimuthal angles. Consequently, during the whole wind bubble evolution, a steady state is never reached and the  expansion of the bow shock   is asymmetric where the regions close to the stagnation expand faster. The final outcome of this process  ($t =$  1.16 Myr)  is a more spherical bow shock than this predicted by the analytical solutions (Eq. \ref{eq:r/ro}), something that is in agreement with what we observe for the case of HFG1. In addition, as the region of the stripped wind material that forms the tail lies away from the wind source, it is not affected by the rapid changes of the wind mass loss rate and thus, it retains its collimated structure formed in the previous stages of the wind bubble evolution.   The final outcome of the wind bubble reveals a cometary structure with an extended bow shock and a collimated tail, something that closely represents the morphology of HFG1.

  \subsubsection{Fast wind and ionization flux description}

 \begin{figure*}
\begin{center}$
\begin{array}{cccc}
&\rm \bf Number~density & \rm \bf Luminosity~per~unit~volume &
\\
\bf D = 310~pc: 
\\
\bf D= 490~pc: &\includegraphics[trim=0 0 0 0,clip=true,width=75.0mm,angle=0]{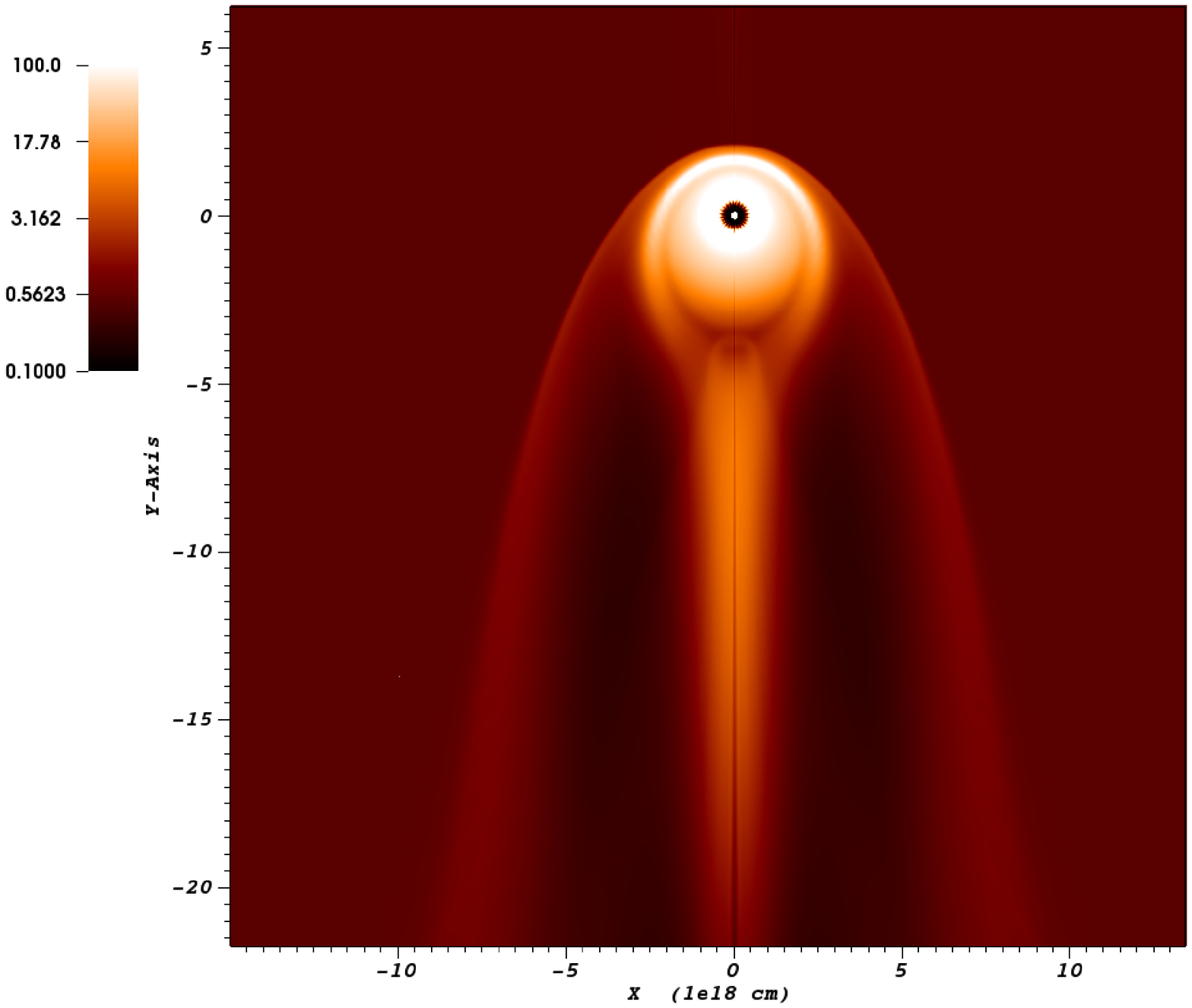} &
\includegraphics[trim=0 0 0 0 clip=true,width=75.0mm,angle=0]{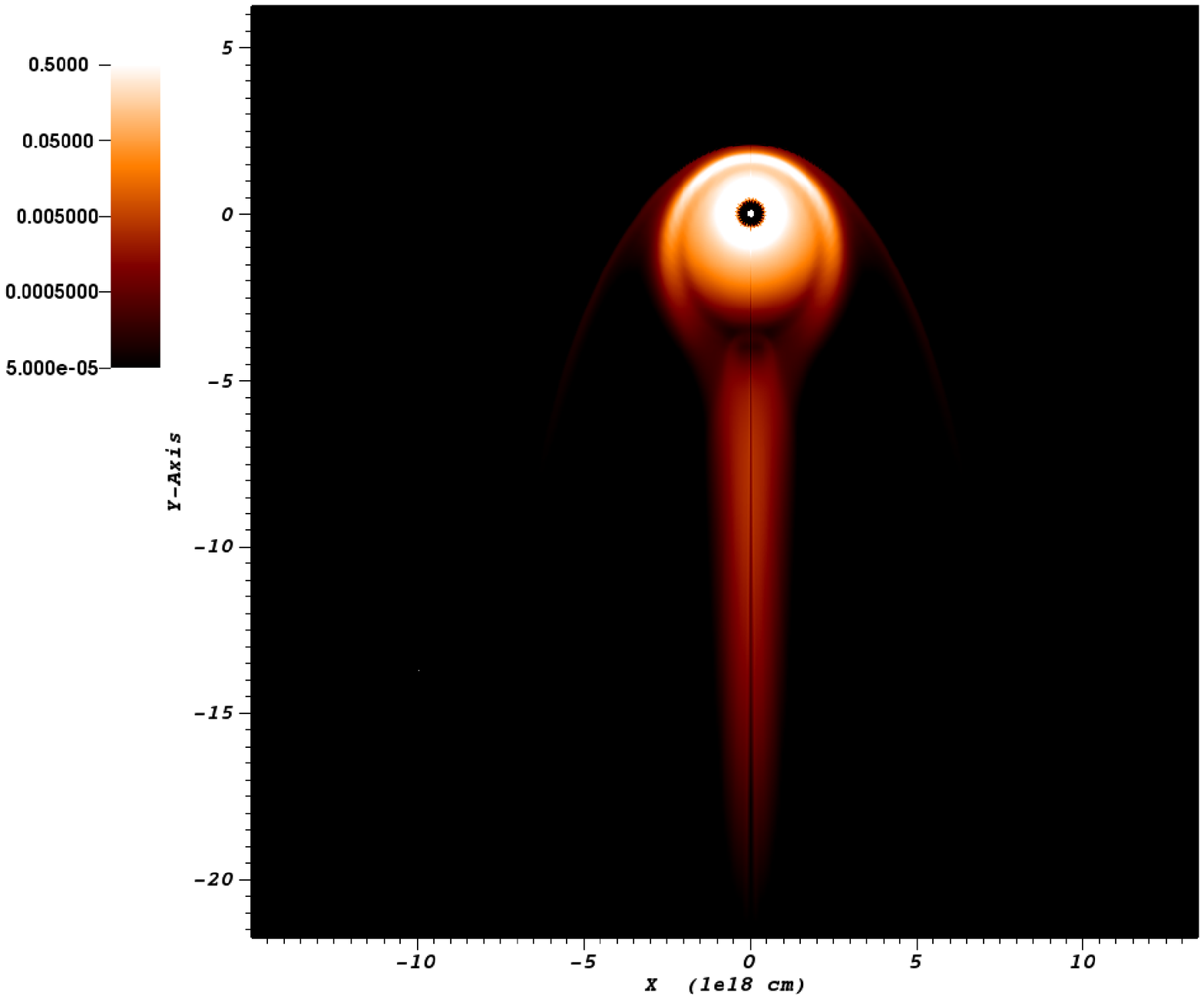}\\
\\
\bf D= 950~pc: &\includegraphics[trim=0 0 0 0,clip=true,width=75.0mm,angle=0]{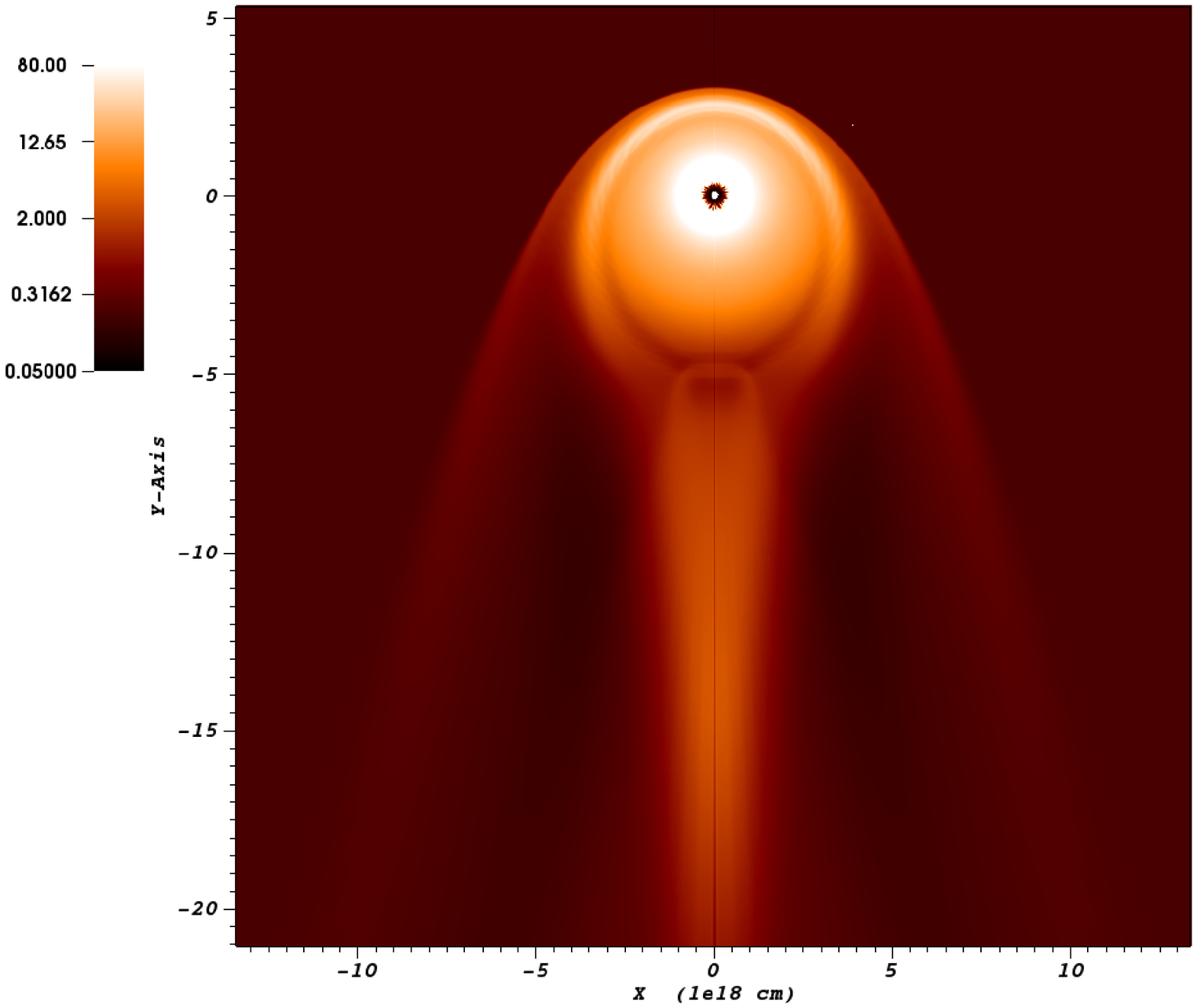} &
\includegraphics[trim=0 0 0 0 clip=true,width=75.0mm,angle=0]{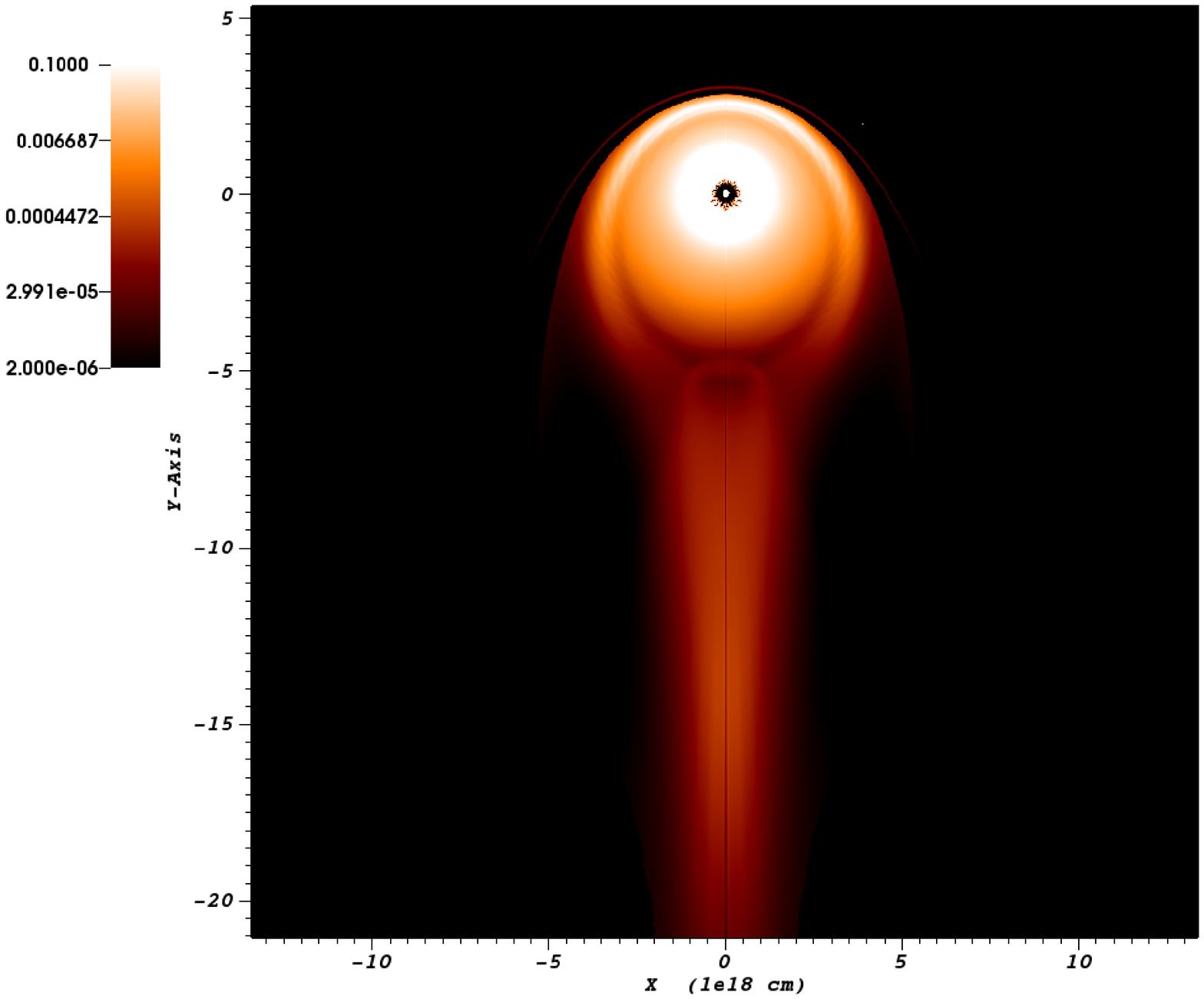}\\
\\
&\includegraphics[trim=0 0 0 0,clip=true,width=75.0mm,angle=0]{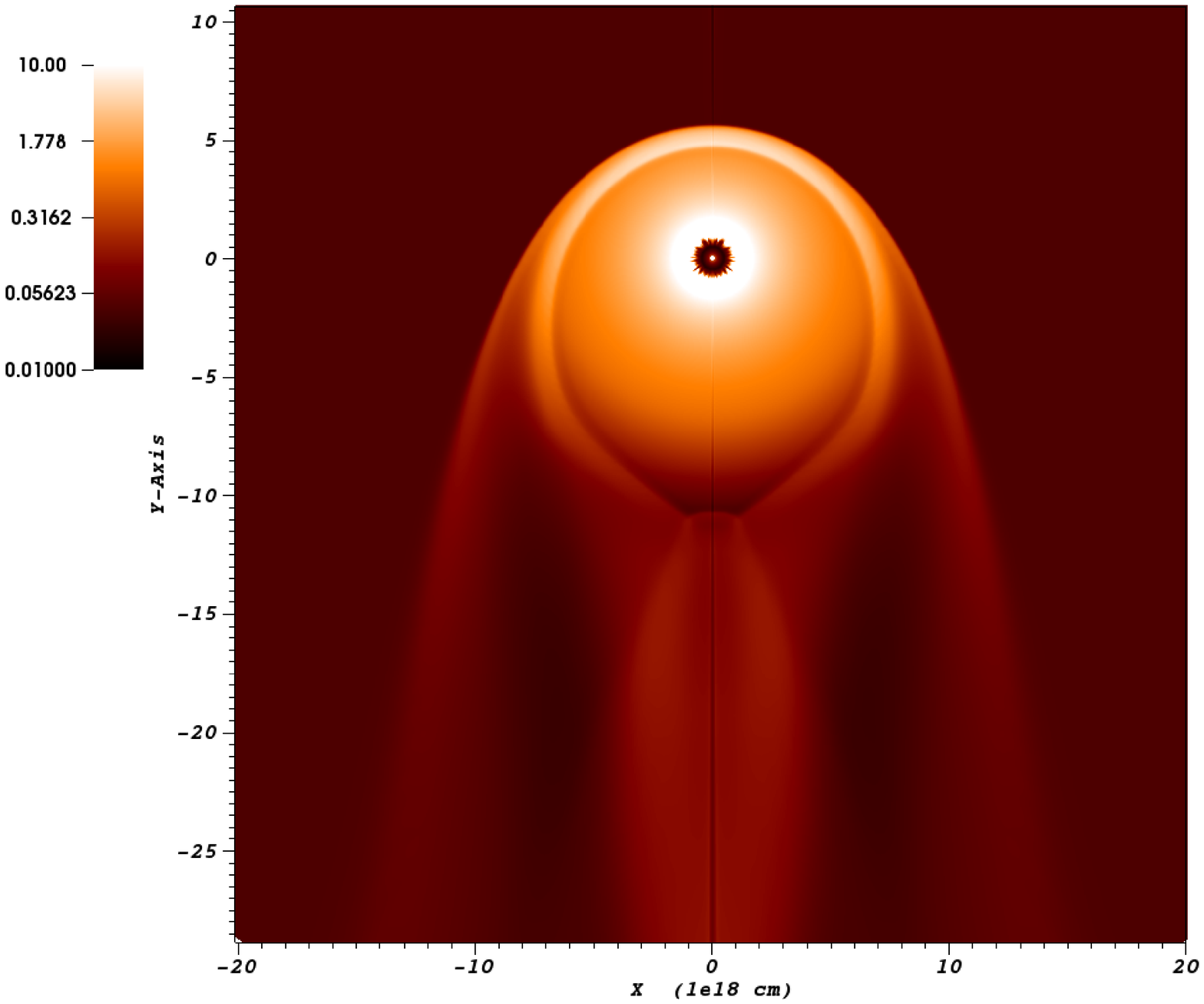} &
\includegraphics[trim=0 0 0 0 clip=true,width=75.0mm,angle=0]{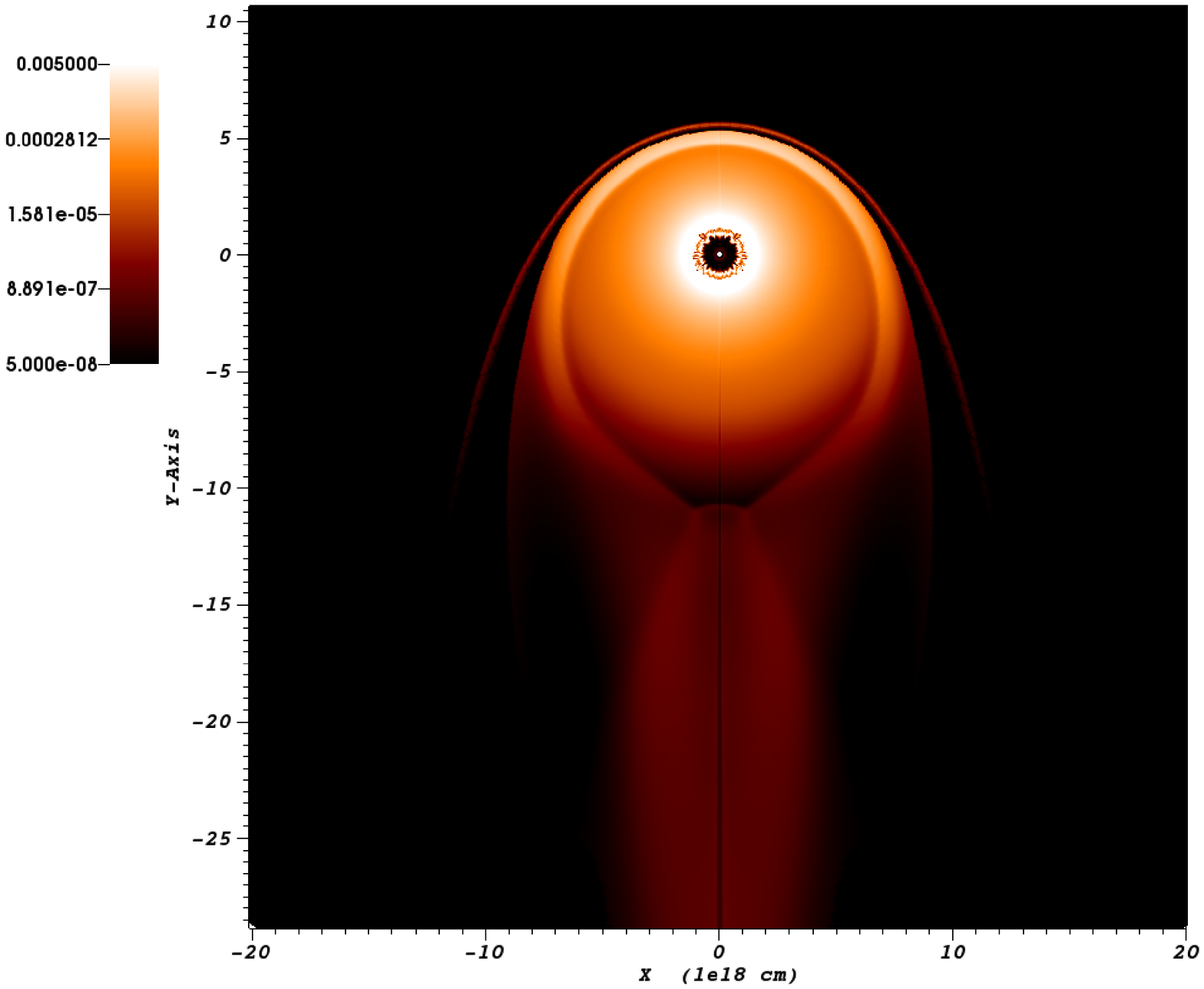}\\
\end{array}$
\end{center}
\caption{Left column: The 2D density profile for the $\rm PN_{3M_{\odot},Dmin}$ (up), $\rm PN_{3M_{\odot},Dopt}$ (middle)  and $\rm PN_{3M_{\odot},Dmax}$ (bottom) models including the fast wind phase. Right column: The luminosity per unit volume for the three models including the ionization effects of the CS (see text for details).   }
  \label{fig:dens_lum}
\end{figure*}

Having achieved to reproduce the main morphological characteristics of HFG1 based on the wind variable models aligned with the AGB evolution of a $3~M_{\odot}$, we proceed our PN modeling by including the short-term phase of the fast wind and the photoinization effects that accompany the collapse of the AGB core.

During the fast wind phase, the wind terminal velocity increases sharply where in about $t \simeq 10^4$~yr  it gets values of the order of $10^3 - 10^4 ~\rm km~s^{-1}$ \citep{Pauldrach88} . To simulate this rapid increase of the wind velocity, we describe the velocity evolution with an hyperbolic tangent function. We demand the initial value of the fast wind velocity to be equal to the AGB wind velocity ($u_{AGB}$) in order to get a smooth evolution from the previous stage. The terminal value of the fast wind velocity is considered to be equal to the current escape velocity  ($u_{esc}$)  of V664 Cas primary star. Given that the observed surface gravity of the sdO central star is $\log \approx 5.65~\rm~cm~s^{-2}$, we find $u_{esc} \simeq 10^3~\rm km~s^{-1}$. Thus, the velocity evolution during the fast wind phase is described by:

\begin{equation}\label{eq:u_fast}
u= 10^{3} \times \tanh \left[ \frac{t}{1.59 \times 10^{-3}} \right] +  u_{AGB}~~\rm km~s^{-1},
\end{equation}
with $t$ the time in Myr. Simultaneously, the mass loss rate decreases in a manner that conserves the wind ram pressure, as follows:

 \begin{equation}\label{eq:dotM_fast}
\dot{M}= \frac{10^{-4.9} \times u_{AGB}} {10^3 \times \tanh  \left[ \frac{t}{1.59 \times 10^{-3}} \right] +  u_{AGB}}~~\rm M_{\odot}~yr^{-1}.
\end{equation}
The above mathematical description holds for $ 7  \times 10^3$~yr, straight after the end of the super wind phase.

 The resulting density structure of the PN for the three studied models, $\rm PN_{3M_{\odot},Dmin}$, $\rm PN_{3M_{\odot},Dmax}$ and $\rm PN_{3M_{\odot},Dopt}$,  including the fast wind phase is depicted at the left column of Fig. \ref{fig:dens_lum}. Due to the small time interval of this phase, the fast wind region remains small compared to the overall PN structure forming a small cavity around the CS without affecting the general morphological properties of the  PN. The small spikes at the  fast wind - AGB interface are resulting from the Rayleigh-Taylor instabilities as the fast wind bubble evolves in the much denser AGB wind region. 
 
 At the final step we attempt to describe in our simulations the ionization front due to the flux of the CS. A numerical approach of this process requires a sophisticated method of radiation transfer something that is not included in the {\sc amrvac} code. Thus, at the current work we resort to a simplified approximation of this process. During the fast wind phase we - instantaneously - increase the cold plasma temperature from values lower than $1000$~K to $10^4$~K. Such a hypothesis is aligned with an R-type ionization front  \citep{Giuliani81,Huarte-Espinosa12}  in which the ionization front evolution is much faster than this of the hydrodynamic evolution. The final profile of PN luminosity per unit volume for the three studied models are portrayed in the right column of   Fig. \ref{fig:dens_lum}.

  In Table \ref{tab:varModels}, we compare the main morphological properties of HFG1 with the relevant values of our modeling.  The models of the time variable wind - contrary to these of constant wind parameters - that correspond to a $3~\rm M_{\odot}$ AGB progenitor  reproduce closely the general shape of the outer shell and the geometrical properties of the tail as observed in HFG1.

\begin{table*}
	\centering
		\caption{Same as Table \ref{tab:Models_comp} but for the three studied wind time variable models.  }

		\begin{tabular}{l |c|c|c|c|c|c| }
		\hline
		\hline
		 & $D~\rm(pc)$& $r_0~\rm(10^{18}~cm)$ & $r_{\theta= 90^o}~\rm(10^{18}~cm)$ & $r_{\theta= 170^o}~\rm(10^{18}~cm)$  & $w_{tail} ~\rm(10^{18}~cm) $  \\
	\hline
Observations                     &310&1.8&2.1&2.6&1.6\\
\hline	
$\rm PN_{3M_{\odot},Dmin}$   &310&1.8&2.1&3.1&1.8\\
\hline
\hline
Observations                     & 490& 2.8 & 3.3 & 4.1 &  2.6\\
\hline	
$\rm PN_{3M_{\odot},Dopt}$   & 490 & 2.8 & 3.2 & 4.7 & 2.8 \\
\hline
\hline
Observations                     &950&5.5&6.3&7.9&4.8\\
\hline	
$\rm PN_{3M_{\odot},Dmax}$   &950&5.3&6.4&9.6&6.1\\
\hline
		\end{tabular}
	\label{tab:varModels}
\end{table*}

 \section{Discussion}\label{sec:Disc}
 
 \subsection{Comparison of HFG1 with the hydrodynamic models}

We have shown that considering time invariant stellar wind and ISM properties, the  morphological properties of HFG1, as revealed by \citet{Heckathorn82} and \citet{Boumis09}, cannot be reproduced. In particular these models cannot reproduce simultaneously the geometry of the  bow shaped shell -which substantially deviates from the steady state solutions (Sect. \ref{sec:bow}) - and the collimated tail observed in HGF1.

Motivated by the stellar evolution theory, we increased by a step the complexity of the modeling by considering a stellar wind with time variant mass loss rates  during the formation of the PN. The description of this time variability  was made by adopting the predictions  
 of AGB \citep{Nanni13} and post-AGB \citep{Pauldrach88} evolutionary models. Finally, the  stellar parameters of the AGB progenitor were determined by comparing the observables of V664~Cas  \citep{Shimanskii04, Exter05} - the central binary of HFG1- with the theoretical post-AGB evolutionary tracks  \citep{Schoenberner83, Bloecker95}.  Within the parameters space that this comparison provided, we tested several models and we found that HFG1 is best reproduced by the mass outflows of a $3 \rm M_{\odot}$ stellar progenitor, with solar metallicity,  during its AGB and post-AGB phase (even though the cases of a 2~\msun~and 4~\msun~AGB  progenitor cannot be strictly excluded).

In our modeling with a 3 \msun~AGB progenitor we took into consideration  three possible distances of the object, two of them correspond to the two distance limits placed by \citet{Exter05}, $D_{min}=$ 310~pc and $D_{max}=$ 950~pc , while the last one to the optimum distance estimated by this work, $D_{opt}=$ 490 pc.  All three models reproduce the overall characteristics of HFG1 outer shell and the collimated trail that follows the PN. According to these models, the tail is formed at the early  TP-AGB phase  where the mass loss rate, and thus the wind ram pressure, is relative low. Nevertheless, as the mass loss rate increases towards super-wind values, the outer shell expands to the observed size. Due to the short time duration of the super-wind phase, the sharp increase of the mass loss rate has not been communicated to the wind material that forms the tail and hence the latter retains its collimated structure. In addition,  as the wind is time variable, the outer shell in our modeling never reaches a steady state and its resulting geometry is more spherical than expected by the analytical approach, something that  resembles closer the observed shell of HFG1.

Except from the general PN morphology that the hydrodynamical simulations and observations reveal, we quantified the comparison between the models and observations using  four parameters (Table \ref{tab:varModels}):  the radius of the stagnation point ($r_o$), the radius of the termination shock  at the region where the angle between the radius vector and the vector of the stellar velocity is $90^o$ and $170^o$ ($r_{\theta= 90^o}$ and $r_{\theta= 170^o}$, respectively) and finally, the width of the tail ($w_{tail}$).  All three models reproduce quite accurately the radius of the stagnation point as well as the radius of the terminations shock at $\theta= 90^o$.  The hydrodynamic simulations reveal the outer shell structure of the PN slightly more elongated than the observed one, resulting to an overestimation of the termination shock radius at $\theta= 170^o$ by around 15 - 20\%. This small difference could possible be explained by the fact the HFG1 is slightly tilted toward the observer \citep{Boumis09} and thus, what we observe  is the protection of the PN in the plane of sky. This, combined with the geometry of HFG1, makes, in this direction,  the observed radius to be smaller than the real one.  Finally, as far as the width of the tail is also well reproduced with the exception of the $\rm PN_{3M_{\odot},Dmax}$ model which overestimates it by  $\sim 25 \%$.

The models also  reproduce  the decrease of the outer shell brightness as the azimuthal angle increases and support the formation  of a `drop-shaped' inner nebula.  Nevertheless, our simulations cannot explain  the sharp positive gradient of  the inner nebula brightness in the direction of PN motion as well as its diffusive appearance. These observed properties could possible be explained by  asymmetric mass outflows during the AGB and post-AGB phase or by a polar angle dependent ionization front. Both processes have not be included in our modeling.

Comparing the three different models, it is clear that the general morphology of the PN does not strongly depend on the fine-tuning of the used parameters. Thus,   the hydrodynamical result cannot draw a firm conclusion about the distance of the object. However, the adaptation of the maximum possible distance of the PN ($D=$ 950~pc) indicates  that HFG1 is one of the largest observed PNe \citep{Frew15} revealing an averaged radius of $\sim 2.1$~pc. In addition, the low AM density adopted in this model ($n_{AM} = 0.04~\rm cm^{-3}$)   is related more to the hot ISM component \citep{Ferriere01}. Considering this, it would be more realistic to adopt an AM temperature of $T \sim 10^6$~K than the used value of $10^3$~K. As we explained in Sect. \ref{sec:Alternatives}, the interaction of the PN with such a high temperature AM would result in a complex, turbulent structure in contrast to the observations. Therefore, even if none of the three models can  strictly be excluded, our results favor more for an HFG1 distance closer to the lower limit of \citet{Exter05}, something that agrees with the independent distance estimation of this study.

 \subsection{SdO in PNe: what can we learn from HFG1}
 
 As we briefly discussed in the Introduction, hot sdO stars are evolved, luminous, low mass stars. Despite decades of research, the origin of these stars remains unclear.  Among other evolutionary paths, it has been suggested that sdO  stars are the remnants of  low mass stars ($M \le 8$ \msun) which have been evolved beyond the AGB phase \citep{Heber09}.  The main argument for this scenario is the position of sdOs in the Herspung-Russel diagram which is consistent with the theoretically predicted post-AGB tracks \citep{Schoenberner83, Bloecker95}. If this scenario is true, it implies that a fraction of sdO  stars should be hosted at the centre of PNe  \citep{Aller13, Aller15}. Thus identifying  sdO stars surrounded by PNe  is an important quest towards a better understanding of the unclear origin of sdO stars.  Up to date a limited number (18) of these  systems have been discovered  \citep[see the list compiled by][and references therein]{Aller15}  and among them is HFG1.   Nevertheless, there is an extra property that makes HFG1 a unique object.
 
 The main component of a PN structure is the bright shell formed by the interaction of the AGB wind with the subsequent fast wind. There are PNe which reveal a more complicated morphology where, in addition to the bright shell, they are consisted of   a low-surface brightness outer rim, the so called `crown' or/and a extended diffusive faint halo. The crown is attributed to the extent of the ionization front resulting to the expansion of the H{\sc ii} region into the neutral undisturbed surrounding. The halo is the remnant of the AGB wind ejected on a previous epoch. Unfortunately, none of these PN structures can provide  direct information about the mass-loss history during the AGB phase. On the one hand, the shell formation is mostly determined by the compact central star properties, and on the other hand it has been shown that development of a multiple component PN structure consisted of a shell crown and halo are independent from the previous AGB-mass-loss histories \citep{Perinotto04}.   
 
 This is not true for the case of HFG1. Its main difference with the general PN structure is that due to the supersonic motion of HFG1 the AGB wind of  its progenitor remains trapped and close to the CS under the ram pressure of the AM. Hence, in this case the outer shell   represents the shocked AGB wind  compressed by the termination shock. The wind properties determine crucially the final morphology of this shell. Thus, both the observed geometrical shape and the size of the outer shell provide important information on the AGB wind properties and its time evolution.

 Our modeling showed that the bow shaped shell of HFG1 can {\it only} be reproduced by a time variable wind which, as time progresses, its mass loss rate increases sharply. This result is aligned with the AGB wind models and their prediction of an AGB super-wind phase. 
 
 Relying our description  on the wind time variability  on the results of AGB evolution models  we find that the best agreement between our simulations and the observed morphology of HFG1 is achieved by assuming an AGB mass of 3 \msun~and solar metallicity.   This is the expected mass for the progenitor of V664~Cas sdO according to the theoretical predictions of post-AGB stars. Thus, our result achieves to  bridge the properties of the PN with the origin and evolution  of its sdO CS. For the first time the predictions of the  theories regarding  the AGB wind properties and  the post-AGB origin for the luminous sdOs are verified based on the geometrical characteristics of the surrounding PN.

 \section{Summary}\label{sec:Sum}

The results of the present work are summarised as follows:

\begin{enumerate}
\item  The main morphological characteristics of the cometary structure of HFG1 can be well explained by the interaction of the local   AM with the mass outflows emanating from its supersonically moving progenitor star, V664~Cas. It has been shown analytically that the properties which characterise AGB winds are able to reproduce the size of the bow shaped outer shell observed in HFG1 by adopting the observed proper motion of the PN and reasonable AM densities.

\item  As the systemic velocity and the size of the PN are distance dependent each model that attempts to explain the properties of HFG1 requires the knowledge of its distance. \citet{Exter05} estimated the distance of HFG1 to be between 310 - 950~pc. Here, we recalculated HFG1 distance by modeling the spectral energy distribution of V664~Cas secondary star. We found that HFG1 is located at a distance of $  D= 490 \pm 50$~pc. 

\item We performed 2D hydrodynamical simulation modeling the interaction of the wind of a supersonically moving AGB star with the local AM. We found that adopting time invariant wind and AM properties, none of our models are able to reproduce the main morphological properties of HFG1. In particular, these models fail to reproduce simultaneously the extended bow shaped outer  shell observed in HFG1 and the collimated trail oriented in the opposite direction of the PN motion. This result is independent from the adopted distance of the object.

\item At the next set of our models we considered that the wind mass loss rate is increasing with time. Such an assumption is in agreement with the stellar evolution theory of AGB star, which predicts that the mass loss rates increase by several order of magnitudes during the AGB evolution.  Based on the observed properties of HFG1 progenitor star, a  $0.57 \pm 0.03$  \msun~sdO star, and the post-AGB evolution theory, 
 we restricted the  mass range of the PN progenitor star  to 1 - 4 \msun. Within this mass range we performed 2D hydrodynamical simulations describing  the wind mass loss rate by using the prediction of TP-AGB evolutionary models.  We found that the morphological properties of HFG1 are best reproduced  by a star with initial mass of 3 \msun~and solar metallicity.

\item Adopting a  3 \msun~progenitor, we modeled HFG1 considering that the PN is placed at the distances of 310~pc, 950~pc \citep[the two distance limits estimated by][]{Exter05}, and 490~pc (the distance as estimated by this work). 
 For all three examined distances,  the hydrodynamic models reproduce closely the overall morphological properties of HFG1. Precisely, they reproduce the geometry of the extended outer shell  explaining its roughly rounded shape which  deviates  from the steady state solution of a bow shaped shell. In addition, a collimated tail in the direction opposite to the PN motion is also formed in these models. Its width and length agree with the observations. Finally, these models reproduce the decrease of the outer shell brightness with the increase of the azimuthal angle as well as the formation of a `drop shaped' inner nebula enclosed on the outer shell.  Nevertheless, our hydrodynamic modeling favors the cases of 310~pc and 490~pc as they result in a more reasonable PN size and AM properties.

\item The rare morphology  of HFG1 combined with our modeling  provides the first  link which connects the theoretical predicted evolution of an AGB star toward the formation of an  sdO with the observed properties of its host PN. This result provides new insights into the unknown nature of sdO stars reinforcing the theory which describes their formation and evolution through AGB and post-AGB phases.

\end{enumerate}

\section*{Acknowledgments}

We thank the anonymous referee for her/his thorough review, which significantly contributed to improving the paper. We are grateful to Rony Keppens for providing us with the {\sc amrvac} code and Ken Crawford for his kind permission to use  the fits files of his HFG1 observations. We thank  Eva Lefa and Carlo Abate for their helpful discussions and their highly appreciable comments. We are thankful also to  John Alikakos for his excellent support during the observations at Helmos Observatory and to  Kosmas Gazeas for providing us the requested time at the Gerostathopoulion Observatory. This work is based on observations made with the ``Aristarchos'' telescope operated on the Helmos Observatory by the Institute of Astronomy, Astrophysics, Space Applications and Remote Sensing of the National Observatory of Athens. A. Chiotellis and N. Nanouris gratefully acknowledge financial support under the ``MAWFC'' project.  Project MAWFC is implemented under the ``ARISTEIA~II" action of the Operational Program ``Education And Lifelong Learning". The project is co-funded by the European Social Fund (ESF) and National Resources.

\bibliographystyle{mn2e}
\bibliography{PNe}


\end{document}